\documentclass[a4paper,11pt]{article}
\usepackage[utf8]{inputenc}
\usepackage[T1]{fontenc}
\usepackage{comment}
\usepackage{tensor}
\usepackage{bbm}
\usepackage{float}
\usepackage{amssymb}
\usepackage{authblk}
\usepackage[dvipsnames]{xcolor}
\usepackage{graphicx}
\usepackage[new]{old-arrows}
\usepackage{jheppub}
\usepackage{bm}

\usepackage[multiple]{footmisc}
\usepackage{multicol}
\usepackage{multirow}
\usepackage{rotating}
\usepackage{array}
\usepackage{mathtools}
\usepackage{longtable}
\usepackage{booktabs}
\usepackage[mathscr]{euscript}
\usepackage{tikz}
\usetikzlibrary{backgrounds}
\usepackage{braket}
\usepackage{subfigure}

\usepackage{dsfont}
\usepackage{slashed}
\usepackage{marginnote}

\newcommand{\RNum}[1]{\uppercase\expandafter{\romannumeral #1\relax}}

\tikzset{
	partial ellipse/.style args={#1:#2:#3}{
		insert path={+ (#1:#3) arc (#1:#2:#3)}
	}
}

\tikzset{snake it/.style={decorate, decoration=snake}}

\usetikzlibrary{calc,decorations.markings}
\usetikzlibrary{decorations.pathmorphing}
\usetikzlibrary{decorations.pathreplacing}
\usetikzlibrary{shapes,backgrounds}
\usetikzlibrary{fadings}
\usetikzlibrary{cd}
\usetikzlibrary{hobby}
\usetikzlibrary{decorations.pathreplacing}
\usetikzlibrary{knots}
\usetikzlibrary{3d}
\tikzfading
[
	name=fade out,
	inner color=transparent!0,
	outer color=transparent!100
]

\newif\ifdraft
\draftfalse

\newcommand{\be}{ \begin{equation}}
\newcommand{\ee}{\end{equation}}
\newcommand{\bi}{ \begin{itemize}}
\newcommand{\ei}{\end{itemize}}

\makeatletter
\newcommand*\bigcdot{\mathpalette\bigcdot@{.65}}
\newcommand*\bigcdot@[2]{\mathbin{\vcenter{\hbox{\scalebox{#2}{$\m@th#1\bullet$}}}}}
\makeatother

\title{\boldmath De Sitter Bra-Ket Wormholes}
\author[a,b]{Alessandro Fumagalli,}
\author[b]{Victor Gorbenko,}
\author[b,c]{Joshua Kames-King,}
\affiliation[a]{Institute for Theoretical Physics, University of Amsterdam, 1090 GL Amsterdam, The Netherlands}
\affiliation[b]{Laboratory for Theoretical Fundamental Physics, Institute of Physics,\\
\'Ecole Polytechnique F\'ed\'erale de Lausanne, Switzerland}
\affiliation[c]{Fields and Strings Laboratory, Institute of Physics,\\
\'Ecole Polytechnique F\'ed\'erale de Lausanne, Switzerland}
\emailAdd{a.fumagalli@uva.nl}
\emailAdd{victor.gorbenko@epfl.ch}
\emailAdd{jvakk@yahoo.com}
\abstract{
We study a model for the initial state of the universe based on a gravitational path integral that includes connected geometries which simultaneously produce bra and ket of the wave function. We argue that a natural object to describe this state is the Wigner distribution, which is a function on a classical phase space obtained by a certain integral transform of the density matrix. We work with Lorentzian de Sitter Jackiw-Teitelboim gravity in which we  find semiclassical saddle-points for pure gravity, as well as when we include  matter components such as a CFT and a classical inflaton field. We also discuss different choices of fixing time reparametrizations. In the regime of large universes our connected geometry dominates over the Hartle-Hawking saddle and gives a distribution that has a meaningful probabilistic interpretation for local observables. It does not, however, give a normalizable probability measure on the entire phase space of the theory. }

\arxivnumber{}

\begin{document}

\maketitle
\flushbottom
\begingroup\allowdisplaybreaks

\section{Introduction}
Understanding the initial conditions for the cosmological evolution of our universe is a fundamental unresolved problem.  One existing proposal relies on the path integral formalism which provides the wave function in the following form:
\begin{equation}\label{eq:formalexpressionwavefunction}
\Psi(\chi_b,h_{i j})= \int^{\chi_b,h_{ij}} \mathcal{D}[\chi]\mathcal{D}[g_{\mu \nu}] e^{i S}\,.
\end{equation}
Here $\chi_b$ and $h_{i j}$ are the ``boundary'' values of matter fields and the metric defined on a given time slice. One of the matter fields or components of the metric can be thought of as time in which case the wave function describes the probability amplitude that the universe occupies a state with the remaining matter fields in the specified configuration and the specified spatial metric. 
Initial conditions are then determined by the class of field configurations over which one integrates. In reference \cite{Hartle:1983ai} a very appealing proposal was made to integrate over smooth complexified geometries. 
Then on inflationary backgrounds dominated by a positive cosmological constant there is a dominant saddle point contributing to \eqref{eq:formalexpressionwavefunction}, that we will refer to as the Hartle-Hawking (HH) state. When the universe is large, we can think of it as approximately classical. This justifies a semi-classical approach, in which one considers quantized matter fields and metric perturbations on a classical background as is done in cosmological perturbation theory. This leads to a density fluctuation spectrum in line with CMB observations \cite{Guth:1980zm,Guth:1982ec}, however the probability distribution for the curvature zero mode appears to be inconsistent with observations: the wave function of the universe calculated with the above proposal implies that a very  short period of inflation is strongly favoured statistically \cite{PhysRevD.37.888}, see \cite{Halliwell:2018ejl,Janssen:2020pii,Lehners:2023yrj,Maldacena:2024uhs} for recent discussions of this problem. 

Nevertheless, the no-boundary proposal is theoretically very appealing: it can be thought of as an effective theory for the initial conditions of the universe which replaces the early time singularity with smooth configurations of fields present in the low energy theory and captures universal properties of the state accessible to a low energy late time observer. 
It stands to reason to look for new contributions to the gravitational path integral, still satisfying the no-boundary  condition, that may in some cases dominate over the HH saddle. Further motivation can be found in the context of black hole physics, in particular black holes in asymptotically anti-de Sitter spacetimes. The situation there is more clear because in addition to a gravitational description there is a more rigorous approach provided by the AdS/CFT correspondence \cite{Maldacena:1997re,Witten:1998qj}. It was recently discovered that apparent inconsistencies on the gravity side are fixed and, in addition, rather subtle features of quantum black hole dynamics are reproduced, once geometries of non-trivial topology are included in the path integral \cite{Penington:2019npb,Almheiri:2019psf,Almheiri:2020cfm,Penington:2019kki}. Many such geometries include connected contributions between universes with two or more asymptotic boundaries.

In inflationary cosmology the boundary is a spatial slice on which the state is defined and in order to compute any physical observable two boundaries are needed: one for the bra and one for the ket of the wave function. Motivated by this idea, \cite{Chen:2020tes} considered geometries that have a connection between the ``universes'' that prepare bra and ket of the wave function, so-called ``bra-ket wormholes'', see figure \ref{fig:densitymatrixcontribution}.\footnote{Similar geometries were previously considered in \cite{PhysRevD.34.2267} and elaborated upon in \cite{Barvinsky:2006uh}. Additional work on bra-ket wormholes can be found in references \cite{Penington:2019kki,Milekhin:2022yzb,Mirbabayi:2023vgl}.} When we compute an observable, for example a correlation function, we couple the degrees of freedom belonging to the two sides. This coupling may have a complicated effect on the observables in the full UV theory, and it can manifest itself as a geometrical connection in an effective theory at early times.  In the context of AdS/CFT, there seems to be increasing evidence that wormholes emerge from a coarse-graining of a more complete microscopic description \cite{Saad:2019lba,Maloney:2020nni,Afkhami-Jeddi:2020ezh,Eberhardt:2021jvj,Kames-King:2023fpa}. Corresponding geometries were identified in \cite{Chen:2020tes} in two-dimensional Jackiw-Teitelboim (JT) gravity coupled to matter, interpreted as cosmology. In fact, in this theory bra-ket wormholes are required purely from theoretical consistency of the state. There is a form of information paradox which arises if only the disconnected HH contribution is included. In spite of their appearance, bra-ket wormholes do not lead to a violation of the purity of the state of the universe, as we discuss below. Thus there still  exists a wave function of the universe, although perhaps a UV completion is necessary for its computation.

In \cite{Chen:2020tes} two models of state preparation were considered: one was based on an EAdS geometry and a bra-ket wormhole in this model is an on-shell solution stabilized by the Casimir energy of matter fields, similar to that in \cite{Maldacena:2018lmt}. On the other hand, no on-shell solution was found in a model based on the de Sitter version of JT gravity. To stabilize the solution a period of Euclidean time evolution had to be added at late times, however, it is not obvious what such a period would correspond to in a more realistic model. 

In this paper we suggest that cosmological bra-ket wormholes can be stabilized in a very natural way if one considers an observable called the Wigner distribution. To understand its meaning let us discuss more closely the boundary conditions we impose on the fields at the late time slice. To be concrete, we focus on dS JT gravity whose bulk action is given by
\cite{Jackiw:1984je,Teitelboim:1983ux,Almheiri:2014cka,Engelsoy:2016xyb,Maldacena:2016upp,Maldacena:2019cbz,Cotler:2019nbi}:
\begin{equation}\label{eq:Jtbulklagrangian}
S_{\text{JT}}= \frac{\phi_0}{4 \pi } \int d^2 x \sqrt{g}R   + \frac{1}{4 \pi }  \int d^2 x \sqrt{g}\phi(R-2)\,.
\end{equation}

We can chose the constant value of the dilaton as time, which we denote by $\phi_b$ and consider the size of the universe $\ell$ as a dynamical variable. This theory has no local degrees of freedom, thus a minisuperspace description is exact. As a remedial object, which does not necessarily correspond to an on-shell solution, let us consider the density matrix for $\ell$:
\begin{equation}\label{eq:formalexpressiondensitymatrix}
\rho(\ell_K,\ell_B|\phi_b)=\int_{\text{no boundary}}^{\phi_b, \ell_B,\ell_K}\mathcal{D}[\phi]\mathcal{D}[g_{ij}] e^{i S}\,.
\end{equation}
 As appropriate for a density matrix, the same Dirichlet boundary conditions are imposed on the time variable, but boundary values of the dynamical variable are different at bra and ket. We now imagine that at large values of $\phi_b$ the variable $\ell$ is semiclassical, and we would like to match it to the classical evolution described by the analogue of the Friedmann equations for this system. However, to specify initial conditions for the classical evolution we do not need $\ell_B$ and $\ell_K$, but rather a unique value of the variable $\ell$ and its time derivative, or more precisely its canonical momentum. Of course, in quantum mechanics it is impossible to simultaneously specify a variable and its momentum. In order to provide such a matching procedure Wigner proposed to take a certain Fourier transform of the density matrix \cite{Wigner:1932eb}:
\begin{equation}\label{eq:formalexpressionwignertransform}
    W(L,P|\phi_b)= \int d\Delta \ell \,\rho\left(L + \frac{\Delta \ell}{2}, L  - \frac{\Delta \ell}{2}\Big|\phi_b\right)e^{ i P \Delta \ell}\,,
\end{equation}
where $W$ is the Wigner distribution.\footnote{See also \cite{PhysRevD.36.3626} for previous work on the Wigner distribution in a quantum cosmology setting.} Clearly it is just the Fourier transform of the density matrix, so for standard normalizable states it contains exactly the same information. For semiclassical states $W(L,P|\phi_b)$ has the meaning of a classical probability distribution on the phase space at time $\phi_b$, which is exactly what we need for the classical equations of motion. In fact, we can simultaneously ``measure'' the universe size and its expansion rate at a given time. At the level of the path integral, the Wigner transform \eqref{eq:formalexpressionwignertransform} changes boundary conditions for the metric and the dilaton and moreover it couples boundary conditions for bra and ket parts of the geometry. As we will see below, this coupling is exactly what is needed to stabilize the wormhole. At the level of equations, it is similar to the double cone wormhole \cite{Saad:2018bqo}. This geometry is the gravitational dual of the ``ramp'' in the spectral form factor for chaotic theories. A genuine bulk solution exists when one performs an integral transform of the partition function to a microcanonical ensemble with respect to the temperature difference.\footnote{ See also \cite{Cotler:2022rud,Chen:2023hra} for developments regarding different couplings on the two sides and matter fluctuations on this background.} This transform is very similar to the Wigner transform \eqref{eq:formalexpressionwignertransform}.

 For classical states we expect the Wigner transform to be dominated by values of $\ell_B$ and $\ell_K$ that are very close to each other, so the resulting geometry is not so different from that considered in \cite{Chen:2020tes}, however, it will now be a semiclassical saddle. In principle, the Wigner distribution should be a well-defined measure on the classical phase space. In our setup we do not expect this to be the case because if $W$ were an integrable function its inverse Fourier transform would also exist and be a smooth function. We know that it is unlikely because we were not able to find a saddle for the density matrix. In the simplest case outlined above $W$ turns out to be a constant, thus its Fourier transform is singular. 

In order to get more interesting distributions we add various matter sources to the theory. In particular, we study conformal matter, a slowly rolling inflaton field, and a combination of both. If the inflaton is used as a time variable, we get a four-dimensional phase space of dilaton, universe size, and their momenta. The distribution has a Gaussian shape that is strongly peaked near classical equations of motion. The distribution is still non-normalizable in the directions that correspond to integration constants in these equations, however, we argue that it still has a meaningful probabilistic interpretation for a local observer who is not sensitive to global properties of the universe, like its total size.

In addition to finding a classical saddle we compute the one-loop determinant around it. This determinant grows with the size of the universe and the value of the dilaton, and, for a large enough universe makes the bra-ket contribution dominate over the disconnected one. This is again similar to the ramp effect in the spectral form factor. 

One of the integration constants in the equations of motion corresponds to the ``age of the universe'', defined as the amount of Lorentzian evolution elapsed since the moment of time when the Hubble parameter, $\dot a/a$, differed from its final value by order one. Ideally we would like to have a distribution that favours old universes. This would cure in our toy model the analogue of the curvature zero mode problem of the higher-dimensional HH state mentioned above. For inflaton matter our distribution is flat with respect to the corresponding parameter, and with the addition of CFT it becomes peaked towards younger universes, thus exhibiting a similar problem. Nevertheless, the problem is milder, and the technical reason for it is different. The  horizon area term which dominates the absolute value of the HH wave function is absent for the wormholes. 

\begin{figure}
    \centering
    \label{fig:densitymatrixcontribution}
    \subfigure[]{ 
    \includegraphics[width=0.15\textwidth]{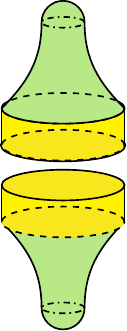}
    }
    \hspace{70pt}
    \subfigure[]{
    \includegraphics[width=0.27\textwidth]{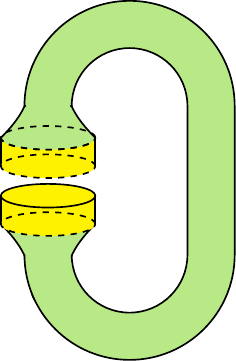}}
    \caption{Here we see two potential contributions to the density matrix as defined in \eqref{eq:formalexpressiondensitymatrix}. 
     The green region denotes a gravitating region, which is glued at the reheating surface to a flat space region (yellow), which we think of as a toy approximation of a weakly gravitating FLRW cosmology. As we are working with JT gravity, only constant positive curvature metrics contribute to the gravitational path integral. (a) The Hartle-Hawking prescription demands that only those complex metrics are considered that are regular and do not exhibit a further boundary in addition to the late time boundary. Note that bra and ket are separately defined objects, such that this is a disconnected contribution to \eqref{eq:formalexpressiondensitymatrix}. (b) A bra-ket wormhole exhibiting a connection in the past. Bra and ket are now connected via a complex contour, which avoids any singularities. The geometries analysed in \cite{Chen:2020tes} are also traced in the future and therefore correspond to a torus.}
\end{figure}

\begin{figure}\label{fig:Wignerfluctuatingboundaries}
    \centering
    \includegraphics[width=0.4\textwidth]{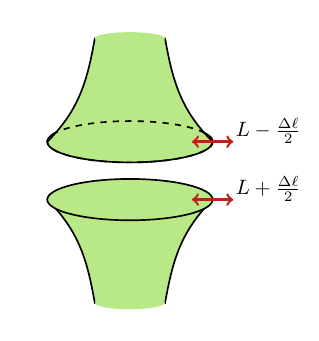}
    \caption{Depiction of the Wigner boundary conditions corresponding to \eqref{eq:formalexpressionwignertransform}. Both bra and ket future infinity are now free to oscillate and the difference is integrated over. Without any further prescription there are disconnected and connected contributions, which correspond to integral transforms of the geometries pictured in figure \ref{fig:densitymatrixcontribution}.}
   \end{figure}

\section{Bra-Ket Wormholes and the Wigner Distribution}\label{sec:braketwhpuregravity}
In the bulk of this paper we will work with dS JT gravity coupled to various matter components. This theory can be thought of as a dimensional reduction of higher-dimensional dS black holes, as explained in \cite{Maldacena:2019cbz}. Thus our discussion should be generalisable also to higher dimensions, although we leave it for future work. 

 The two-dimensional action with boundary terms corresponding to Dirichlet boundary conditions reads \cite{Jackiw:1984je,Teitelboim:1983ux,Almheiri:2014cka,Engelsoy:2016xyb,Maldacena:2016upp,Maldacena:2019cbz,Cotler:2019nbi}:
\begin{equation}\label{eq:actionfordirichlet}
S_{\text{DD}}= \frac{\phi_0}{4 \pi } \left(\int d^2x \sqrt{g}R +2 \int du \sqrt{h} K \right) + \frac{1}{4 \pi }  \int d^2x \sqrt{g}\phi(R-2)\, + \frac{\phi_b}{2 \pi} \int d u \sqrt{h} K\,,
\end{equation}
where $h$ refers to the induced metric and $K$ to the corresponding extrinsic curvature. The first term is purely topological and amounts to the standard de Sitter entropy. The coordinate $u$ runs over the wiggly boundary of JT gravity, and reduces to the spatial coordinate of the bulk metric when minisuperspace approximation is used. We will consider such a limit up until section \ref{sec:oneloopdeterminant}, where we start including the effects of the boundary mode. The equations of motion with respect to the dilaton constrain the metric to be $dS_2$. The equations of motion with respect to the metric instead imply that $\epsilon^{\mu \nu} \partial_{\nu}\phi$ constitutes a Killing vector. 
There are two types of basic solutions we will discuss. First we have the global dS solution:
\begin{equation}\label{eq:globalmetric}
\renewcommand{\arraystretch}{2}
\begin{array}{llr}
ds^2 = - d\hat{t}^2 + \cosh(\hat{t})^2 \, d\hat{\varphi}^2 \,, & \quad \phi = A_G \sinh(\hat{t})\,, &\\
ds^2 = \dfrac{-d\hat{\eta}^2 + d\hat{\varphi}^2}{\sin(\hat{\eta})^2} \,, &\quad \phi = \dfrac{A_G}{-\tan(\hat{\eta})} \,, &\qquad \hat{\varphi} \sim \hat{\varphi} + 2\pi\,.
\end{array}
\end{equation}
These are the standard coordinates used in constructing the Hartle-Hawking geometry. To construct connected solutions, we will consider the Milne solution:
\begin{equation}\label{eq:milnemetric}
\renewcommand{\arraystretch}{2}
\begin{array}{llr}
ds^2 = - dt^2 + \sinh(t)^2 \, d\varphi^2 \,, &\quad   \phi = A \cosh(t)\,, &\\
 ds^2 = \dfrac{-d\eta^2 + d\varphi^2}{\sinh(\eta)^2} \,, & \quad \phi = \dfrac{A}{-\tanh(\eta)}\,, & \\
ds^2 = \dfrac{-dv^2}{v^2 - 1} + (v^2 - 1) \, d\varphi^2 \,, & \quad \phi = A \, v\,, & \qquad \varphi \sim \varphi + 2\pi b\,.
\end{array}
\end{equation}
This is the coordinate system we use in the construction of our bra-ket solutions. Both coordinate systems exhibit exponential asymptotic growth of the dilaton at the future boundary, where it approaches a value $\phi=\phi_b$. We would like to think of this as the reheating surface. In particular, we can consider attaching a non-gravitating flat Minkowski cylinder to it, as in figure \ref{fig:densitymatrixcontribution}, which is an approximation of a weakly gravitating FLRW evolution. Note that periodicity of the spatial coordinate is fixed in the global case in order to avoid the singularity at the pole of the half-sphere $t+i \pi/2$, according to the no-boundary prescription. Instead, \eqref{eq:milnemetric} has arbitrary periodicity, parametrized by $b$ and a  singularity at $t=0$.

\subsection{Bra-ket wormholes in pure gravity}\label{sec:bra-ketwormholes}
Let us now come to finding saddles of the path integral that computes the Wigner distribution defined in \eqref{eq:formalexpressionwignertransform}. While naturally the Hartle-Hawking contour constitutes a disconnected contribution as bra and ket are separately defined quantities, we are looking for spacetimes that connect bra and ket in the past. The main idea is to consider the basic Lorentzian de Sitter solution in the Milne patch \eqref{eq:milnemetric} and let the contour of the time coordinate reach a second boundary. The contour deviates into the Euclidean direction as to avoid the singularity of the Milne coordinates, such that bra and ket are connected through a complexified smooth geometry. One can think of this geometry as consisting of three regions: a Euclidean region that creates a thermal state for the matter fields and two Lorentzian regions that evolve the bra and the ket correspondingly. 

Three such contours were discussed in reference \cite{Chen:2020tes}, which differ by the overall Euclidean displacement between bra and ket endpoints.\footnote{Contours that can be deformed into each other without crossing the singularities are equivalent.} This parameter (see eq. \eqref{eq:temperature} in the next section), denoted by $\beta$ therefore fixes the temperature of (conformal) matter once it exits the Euclidean region. Here we will focus on the the so-called $\pi$-contour, see figure \ref{fig:Wignerpicontour}, as it is the simplest contour that has a finite temperature. Both the metric and the dilaton grow asymptotically at the two boundaries.  

In \cite{Chen:2020tes} it was found that there is no  classical saddle point with Dirichlet boundary conditions corresponding to a density matrix. This is also the case for other boundary conditions imposed individually on bra and ket parts, for example Neumann. They simply correspond to computing a density matrix but in a different representation. Let us quickly recap this discussion: if we consider the Milne patch in FLRW coordinates (coordinate system \eqref{eq:milnemetric}) then a general density matrix element amounts to the following boundary conditions 
\begin{align}
    b \sinh(t_K) &=\ell_K\,,\nonumber\\
    b \sinh(t_B) &=\ell_B\,,\\
    A \cosh(t_B) &= A \cosh(t_K)= \phi_b,\nonumber
\end{align}
where the dilaton plays the role of the clock field. The dilaton condition is only solved if the contour endpoints are of the form: $t_B= t_K \pm 2\pi i$. This implies that metric conditions become degenerate, and the solution exists only for $\ell_K=\ell_B$, that is for diagonal elements of the density matrix. For those the $b$ mode remains unfixed, thus there is no unique classical solution either. Therefore the density matrix is of the form $\rho \sim \delta(\ell_K-\ell_B)$, since the divergent integral over the $b$ mode produces the singularity. 

The delta-function corresponds to  a density matrix that is maximally mixed. Recently, maximally mixed states were discussed in the context of de Sitter physics \cite{Lewkowycz:2019xse,Lin:2022nss,Witten:2023xze,Milekhin:2024vbb}, and it would be interesting to see if the delta-functional form of the density matrix that we see has any relation to this discussion.

The Wigner distribution \eqref{eq:formalexpressionwignertransform} admits a path integral representation with a different set of boundary conditions compared to a density matrix.{\footnote{We will use lower-case letters for the density matrix variables and capital letters for the variables of the Wigner distribution. The time variable will be denoted by a lower-case letter with lower index $b$. The same index will be used when a diagonal element of the density matrix is considered. More details on the definition and properties of the Wigner transform can be found in Appendix \ref{sec:app_wigner}.}} In particular, we fix the sum of the boundary lengths $L =\frac{\ell_K+ \ell_B}{2}$, and the conjugate momentum of the difference of the boundary lengths $\Delta \ell=\ell_K-\ell_B$, which as we show in  Appendix \ref{secapp:gravitationalaction}, turns out to be a sum of the normal derivatives of the dilaton at the two boundaries $P=-\frac{(\partial_n\phi_K+\partial_n \phi_B)}{2}$ (the minus sign is just for convenience). This implies that the contour is not necessarily closed, in a sense that the value of the metric can be different at the two endpoints. This is most transparent in the $v$-variable, see figure  \ref{fig:Wignerpicontour}c in which the metric is not a periodic function.
On the $\pi$-contour and its complex deformations, the length of the ket and bra boundaries are defined as 
\begin{equation}\label{eq:boundarylengthdefinition}
\ell_{K,B}= \frac{1}{2 \pi} \int d \varphi \sqrt{h}_{K,B}\,,
\end{equation}
where $\sqrt{h}$ is the induced volume form at the boundary and the branch of the square root is chosen in such a way that the real parts of both $\ell_K,\ell_B$ are positive. The factor of $1/2 \pi$ is added for later convenience.
As a first step we must determine the required boundary terms that ensure a well defined variational principle. This is explicitly done in Appendix \ref{secapp:gravitationalaction}. In short, we must add an additional boundary term compared to the standard action \eqref{eq:actionfordirichlet}. We have to use 
\begin{equation}\label{eq:boundaryactionWigner}
S_{W}=\frac{1}{2 \pi} \int d \varphi (\Delta \sqrt{h} ) P + S_{\text{DD}}= \Delta \ell \,P + S_{\text{DD}}\,.
\end{equation}
The additional term is the analogue of the term in the exponent in \eqref{eq:formalexpressionwignertransform}.
Note that we are still using standard 
Dirichlet conditions on the dilaton. This is motivated by the idea that we would like to think of it as the clock field in the current setting. 
We will in later sections also consider the introduction of an inflaton, which may also function in a similar way.

Now that we have specified the Lagrangian, equation \eqref{eq:boundaryactionWigner}, implied by our boundary conditions, we can start searching for solutions. 
\begin{figure}
    \centering
    
    \subfigure[]{ 
\includegraphics[width=0.25\textwidth]{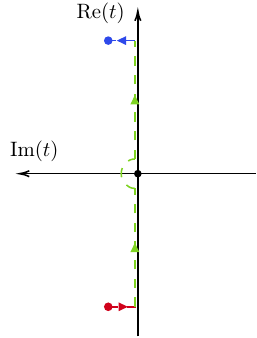}
}
    \hfill
    \subfigure[]{
\includegraphics[width=0.25\textwidth]{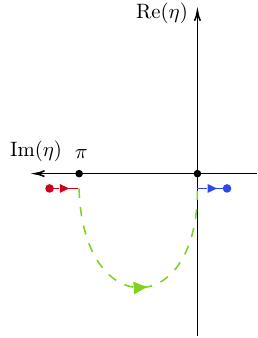}
}
      \hfill
    \subfigure[]{
\includegraphics[width=0.25\textwidth]{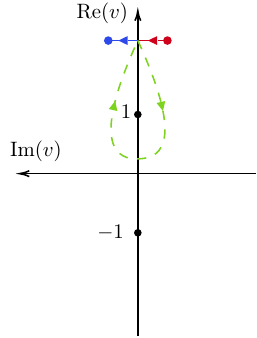}
}
    \caption{The perturbed $\pi$-contour in FLRW coordinates a) and conformal coordinates b) and $v$ coordinates c). The original contour analysed in \cite{Chen:2020tes} is depicted in green dashed lines, deviations are depicted as blue and red solid lines. The bra and ket endpoints of the new contour are depicted as red and blue dots respectively. The singularities in the metric are denoted by black dots. The new set of boundary conditions used for the Wigner distribution allow for complex shifts of the contour endpoints.}
    \label{fig:Wignerpicontour}
\end{figure}
We will be working with the Milne background \eqref{eq:milnemetric} on the contour in figure \ref{fig:Wignerpicontour} and include arbitrary displacements in the Euclidean direction at the endpoints.\footnote{Our contour violates the criterion proposed in \cite{Witten:2021nzp,Kontsevich:2021dmb}. Nevertheless we do not observe any instabilities on our solutions. It would be interesting to understand in detail what the implications of this violation are for our setup. Recently the criteria in application to the HH wave functions were discussed in \cite{PhysRevLett.131.191501,Maldacena:2024uhs,Hertog:2024nbh,Lehners:2023pcn,Janssen:2024vjn,Jonas:2022uqb}.} The properties of the Wigner distribution impose some constraints on the displacements: we need to ask that the arguments $L,P$ are real, since for example $L$ is the physical parameter that corresponds to the size of the universe. This requirement forces the following parametrization (in FLRW coordinates) 
\begin{align}\label{eq:braketcontoursFLRW}
t_K&=t_c + i \,t_E\,,\\
t_B&=-t_c + i \,t_E\,\nonumber,
\end{align}
where $t_c>0$ and $t_E \in \mathbb{R}$.\footnote{There is another contour which instead has $t_c<0$ and corresponds to a contracting universe. We will not consider this branch and assume that a projection was made on an expanding universe.} This ansatz for the contour will in general lead to complex lengths of the individual bra, ket boundaries but importantly the sum remains real as dictated by our boundary conditions. Hence, the parameter $t_E$ captures how far the geometry deviates at late times from standard de Sitter asymptotics or equivalently how discontinuous the geometry is. We will see that for the vacuum configuration we are considering here, $t_E$ is set to zero (dynamically), whereas by inclusion of backreaction from the matter fields it acquires a non-zero value. The parameter $t_c$ instead governs the number of e-folds in this model, i.e. if the universe is ``old'' or equivalently if the boundary is at late times. This is also the quantity that must be large to consider the usual Schwarzian limit of JT gravity. In the later sections we will often use expansions with respect to this parameter.

The contour ansatz \eqref{eq:braketcontoursFLRW} leads to the following boundary condition for the metric in the Milne coordinate system \eqref{eq:milnemetric}:
\begin{align}\label{eq:metricboundaryconditionlbar}
L =  b \cos(t_E) \sinh(t_c)\,.
\end{align}
For any on-shell solution the parameters $b$, $t_E$ and $t_c$ should be fixed dynamically such that the overall geometry is determined uniquely. Before considering any possible addition of matter let us first determine the vacuum solution. It is often easier to not solve the dilaton equations directly but only the metric sector and then consider an off-shell action and extremise with respect to the remaining parameters $b$ (or $t_c$) and $t_E$. This is equivalent to fixing those parameters via the boundary conditions in the dilaton sector.
The action \eqref{eq:boundaryactionWigner} in FLRW coordinates is given in \eqref{eq:FLRWactionW(L,P,phib)}. The contour ansatz \eqref{eq:braketcontoursFLRW} then leads to the following off-shell expression:
\begin{align}\label{eq:gravitycontribution}
i S_{W(L,P|\phi_b)}= 2\,b \sin(t_E)\left(\sinh(t_c)\phi_b-\cosh(t_c)P \right)\,.
\end{align}
We can express $t_c$ via $b$ by use of the boundary condition \eqref{eq:metricboundaryconditionlbar} to arrive at an off-shell action 
\begin{equation}\label{eq:gravitycontributionlatetimeexpanded}
    i S_{W(L,P|\phi_b)}=2 L \phi _b \tan \left(t_E\right)- 2 P \sin \left(t_E\right) \sqrt{b^2+\frac{L^2 }{\cos ^2\left(t_E\right)}}\,.
\end{equation}
Note the $2\pi$ periodicity of the action with respect to $t_E$. Extremising this action with respect to $b$ and $t_E$, leads to the following on-shell values:
\begin{align}\label{eq:onshellvalues vacuum}
t_E&=0\,,\\
b&= \frac{L \sqrt{ \phi_b^2 - P^2}}{ P}\nonumber\,.
\end{align}
We assume all phase space variables $L , P, \phi_b$ to be positive and large. There are two branches for $b$ and here we have taken the branch in line with $b>0$. We can see that reality of $b$ is guaranteed for the range $\phi_b \geq P >0$.\footnote{We include the point $b=0$ as in a path integral treatment of JT gravity this is integrated over. Presumably in the presence of matter a UV completion is required.} Hence there is a restriction on phase space. The solution \eqref{eq:onshellvalues vacuum} is a saddlepoint.
 Let us emphasise that the action is stationary with respect to both parameters such that this is an honest bra-ket solution of the JT theory.  Therefore, the problem of the unfixed mode present in \cite{Chen:2020tes} is circumvented. The physical difference is that the new Wigner boundary conditions couple the two boundaries in a non-trivial way, stabilizing the wormhole. The action on-shell is  zero as the imaginary shift at the endpoints is set to zero. This may therefore seem like a flat Wigner distribution. However, it only exists for the aforementioned range $\phi_b \geq P > 0$, i.e there is a restriction on the region of phase space. In addition we should also note that we have only shown that the exponential of the vacuum solution is on-shell zero. We should expect loop effects to contribute a prefactor. So we include a normalisation factor, such that we have 
\begin{equation}\label{eq:vacuumLPphibWignerfuncton}
    W( L,P|\phi_b)=\mathcal{N}_{L,P}\,,
\end{equation}
where $\mathcal{N}_{L,P}$ generally may be a function of $(L,P ,\phi_b)$. We will determine this function in section \ref{sec:oneloopdeterminant}. 

As mentioned above, there are essentially two equivalent ways of solving the system.\footnote{In section \ref{sec:oneloopdeterminant} we actually use a third approach, in which we first consider an off-shell density matrix and integrate both over the internal modulus $b$ and perform the appropriate Wigner transform.} Here we have taken the approach of only considering the metric boundary conditions and then extremising with respect to the remaining free parameters. It is instructive to outline also the other approach. For this we first solve the equations of motion for the dilaton, which for the vacuum case at hand brings us back to \eqref{eq:milnemetric}. Then we solve for the integration constants $b$, $A$, $t_c$ and $t_E$ in such a way that the boundary conditions are respected. These on-shell values for the bulk parameters are then inserted into the action. So in addition to the metric sector boundary condition \eqref{eq:metricboundaryconditionlbar}, we have to consider the dilaton sector. Clearly the Dirichlet conditions on the boundary value $\phi_b$ can only be solved for $t_E=0$, such that the dilaton boundary conditions reduce to 
\begin{align}\label{eq:dilatonvacuumvalues}
P&= A  \sinh(t_c)\,,\\
\phi_b &= A \cosh(t_c )\nonumber\,.
\end{align}
Solving this system of equations together with \eqref{eq:metricboundaryconditionlbar}(for $t_E$ equals zero) leaves us with on-shell values \eqref{eq:onshellvalues vacuum} and in addition the following value for the dilaton integration constant
\begin{equation}
    A=\sqrt{\phi_b^2-P^2}\,.
\end{equation}

It is also instructive to have a closer look at the real value of the coordinate time at the contour endpoint: 
  \begin{equation}
 t_c=\log\left(\frac{\sqrt{ \phi_b  + P}}{\sqrt{ \phi_b-P}}\right)\,.
 \end{equation}
The Wigner distribution we study now is not a function of $t_c$, but of $L$, $P$ and  $\phi_b$.
We can see that old universes (large $t_c$) correspond to the asymptotic approach $P \rightarrow \phi_b$, while large or small $L$ is not correlated with the age of the universe. This is different than in the HH case because there is no geometric constraint on the size of the universe when it exits the Euclidean region, such that it can be arbitrarily large or small. The relation between the dilaton and its time derivative instead controls the length of Lorentzian evolution. It is still meaningful to ask whether our current distribution prefers old or young universes. The answer is that it is flat with respect to this parameter as well.

 Let us also note that both extremising and solving the boundary conditions explicitly constrains us to the unperturbed $\pi$-contour of \cite{Chen:2020tes}, which now is a solution of the theory. Of course we can also determine a Wigner function of the type $W(L,P|\phi_b)$ for the Hartle-Hawking proposal by taking the on-shell bra and ket wave functions and performing the Wigner transform of the on-shell result. This is done in Appendix \ref{app:wignertransformofHH}.
\begin{figure}
    \centering
    \includegraphics[width=0.35\textwidth]{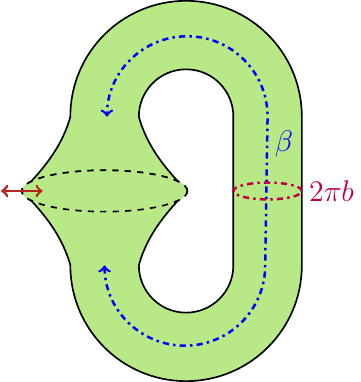}
    \caption{Illustration of the bra-ket wormhole with Wigner boundary conditions and its geometric parameters. The two cyles of the torus are depicted as dashed, coloured lines. The blue cycle along the time direction sets the temperature for matter fields and corresponds to the Euclidean difference between bra and ket. While for the basic contour of \cite{Chen:2020tes} it is set to $\pi$ it can receive corrections with the new boundary conditions. The second cycle exhibits a minimal length geodesic of size $2 \pi b$ but grows with the expansion of the universe towards the two boundaries. The asymptotic bra and ket endpoints of the contour may now fluctuate, which is depicted as a red double-arrow.}
    \label{fig:braketwormholewithparameters}
\end{figure}

\section{Addition of CFT Matter}\label{sec:braketwhandconformalmatter}
It is interesting to see how the Wigner distribution changes in the presence of matter. We start with conformal matter coupled to the metric, such that the stress tensor will backreact on the dilaton. The addition of matter may in principle lift $t_E$ to non-zero values and hence ``open up'' the contour. We should therefore think of this Euclidean shift as a quantum effect. In a cosmological context conformal matter models radiation which is more important at early times and dilutes as the expansion proceeds. Our approach is, as usual, to integrate out the CFT degrees of freedom to arrive at an effective action for the CFT which is just the logarithm of the partition function on the torus
\begin{equation}
    iS_\text{CFT}=\log(Z_\text{CFT})\,.
\end{equation}
We consider periodic boundary conditions for the CFT on the time contour, which corresponds to taking the trace over the matter fields.  Then the total Wigner distribution is
\begin{equation}
W_{\text{CFT}}(L,P|\phi_b)= \int_{L,P,\phi_b} \mathcal{D}[\phi]\mathcal{D}[g_{\mu \nu}] e^{iS_{W(L,P;\phi_b)}} Z_\text{CFT}\,,
\end{equation}
where the gravitational action $S_{W(L,P|\phi_b)}$ is \eqref{eq:boundaryactionWigner} with the appropriate boundary conditions. 
As we are considering the addition of matter the global properties of the geometry, shown in figure  \ref{fig:braketwormholewithparameters}, play a decisive role for the matter contribution, since the partition function depends on the two cycles of the torus $b$ and $\beta$. On the contour ansatz \eqref{eq:braketcontoursFLRW} the former reads 
\begin{align}\label{eq:temperature}
\beta &= i\int_{t_B}^{t_k}\frac{dt}{\sinh(t)}\nonumber\\
 &=i \log(\tanh(t/2))|_{t_B}^{t_K}\nonumber\\
 &=\pi - 4 e^{-t_c} \sin(t_E)+ ...  \,,
\end{align}
where we have performed an expansion with respect to the large parameter $t_c$, corresponding to a universe that is old. Both the stress tensor and the partition function are affected by the positively curved spacetime i.e a conformal anomaly is induced, which gets added to the flat space result. Overall, the structure we assume for the CFT contribution is 
\begin{equation}\label{eq:splitupcftpartitionfunction}
\log(Z_{\text{CFT}})= \log(Z_{\text{flat}}) + \log(Z_{\text{an}})\,.
\end{equation} 
Even though the full geometry is in general discontinuous at the future bra-ket junction, the geometry without the conformal factor is still continuous, so the calculation of the flat space partition function follows the standard procedure. Discontinuities of the geometry will show up in the conformal anomaly contribution.  Here we only the consider the propagation of the vacuum mode along the spatial cycle, which is a valid approximation for $b\gg1$:
\begin{equation}\label{eq:flatspacecftcontirbution}
    \log Z_\text{flat}= \frac{c \pi (2 \pi b )}{6 \beta} = \frac{c}{3} b \frac{\pi^2}{\beta}\,.
\end{equation}
The conformal anomaly for an Euclidean metric written in the form $ds^2= e^{2\omega} d \hat{s}^2$ reads
\begin{equation}
    \log Z_{\text{an}}=  \frac{c}{24 \pi} \int d^2x \sqrt{\hat{g}}\left[\hat{R} \omega+ (\hat{\nabla}\omega)^2\right]+  \int_\partial d \varphi \sqrt{\hat{h}}\hat{K} \omega\,.
\end{equation}
By Wick rotating the Lorentzian Milne metric \eqref{eq:milnemetric} $\tau=i \eta$ we can rewrite the anomaly contribution as
\begin{align}
    \log Z_{\text{an}}=-\frac{c  b \beta}{12 }  -i\frac{c}{24 \pi} \int d^2x \sqrt{-g}\,,
\end{align}
where $\beta=i \Delta \eta$ and the quantities refer to the Lorentzian metric \eqref{eq:milnemetric}. The last term is proportional to the volume form and can be reabsorbed in the definition of $\phi_0$.\footnote{The same term appears when including a CFT matter in the Hartle-Hawking state so that redefining $\phi_0$ removes both terms.} This term is also the only one that is sensitive to the geometry being discontinuous.
The effective CFT action as a function of the mode $b$ and the temperature \eqref{eq:temperature} is:
\begin{equation}\label{eq:CFTpartitionfunction}
\log Z_{\text{CFT}}= \frac{b c}{3}\left(\frac{\pi^2}{\beta}-\frac{\beta}{4}\right)\,.
\end{equation}
On our contour, expanding at late times, it becomes
\begin{equation}\label{eq:cftcasimircontribution}
\log Z_{\text{CFT}}= \frac{b \pi c}{4} + \mathcal{O}(e^{- t_c}) \,.
\end{equation}
This contribution now must be added to the gravitational action \eqref{eq:gravitycontributionlatetimeexpanded} which expanded at late times reads
\begin{align}\label{eq:actiongravitylatetimes}
    iS_{W(L,P|\phi_b)}=2 L \left(P - \phi _b\right) \tan \left(t_E\right)-\frac{b^2 P \sin \left(t_E\right) \cos \left(t_E\right)}{ L}+ \mathcal{O}(e^{- t_c}).
\end{align}
Note that the CFT terms also respect the $2\pi$ periodicity of the gravitational contribution. So that the total action is
\begin{align}
    i S_{W(L,P|\phi_b)}^{\text{CFT}}=iS_{W(L,P|\phi_b)}+\log(Z_\text{CFT})\,.
\end{align}
We again extremise with respect to $t_E$ and $b$. We only consider the branch which furnishes positive and real on-shell values. These are of the form 
\begin{align}\label{eq:o1saddles}
    b^*&= \frac{ L \left(4 (\phi _b^2-P^2)+\frac{\pi ^2 c^2}{8}\right)}{\left( \phi _b+ P\right) \sqrt{4 (\phi _b^2 -P^2) + \frac{\pi ^2 c^2}{16}}}\,,\\
    t_E^*&= \arctan\left(\frac{\pi  c}{\sqrt{64 \left( \phi _b^2-P^2\right)+\pi ^2 c^2}}\right)\,.
\end{align}
Naturally these expressions agree with the vacuum solution in the $c$ to zero limit.\footnote{Note that $P=\phi_b + \mathcal{O}(1/\phi_b)$ at late times.} We see that $t_E$ is set to a non-zero value as long as $c$ is non-zero. The imaginary shift of the contour endpoints therefore has been lifted to some finite value in the presence of the CFT. Therefore the action now also takes on a non-zero value, namely
\begin{align}\label{eq:CFTonshellOrder1}
    i S_{W(L,P|\phi_b)}^{\text{CFT}}=\frac{\pi  c L  \sqrt{4 (\phi_b^2-P^2)+\frac{\pi ^2 c^2}{16}}}{ \left( \phi _b+P\right)}\,.
\end{align}
Hence the distribution (up to corrections of order $e^{-t_c}$ that we drop) is 
\begin{equation}\label{eq:CFTWignerdistribution}
    W_{\text{CFT}}(L,P|\phi_b)= \mathcal{N}_{ L,P}\text{exp}\left(\frac{\pi  c L  \sqrt{4 (\phi_b^2-P^2)+\frac{\pi ^2 c^2}{16}}}{ \left( \phi _b+P\right)}\right)\,,
\end{equation}
where we have again included a normalisation factor.
These expressions are exact in $c$ but expanded at late times. Again, this is a genuine solution of the theory in which all parameters have been solved. By looking at the off-shell action as a function of $t_E$ and $b$, we see that this solution is a saddle point, in particular it is a maximum in the $b$ direction and a minimum in the $t_E$ one. It is also instructive to consider the on-shell value of $t_c$:
\begin{align}\label{eq:tconshellCFT}
    e^{t_c}=\frac{2 (\phi _b+P)}{ \sqrt{4 (\phi _b^2-P^2)+\frac{\pi ^2 c^2}{8}}}\,.
\end{align}
Reality (and positivity) of all the above parameters and the action is guaranteed if we restrict $P$ to the range: $\sqrt{\frac{\pi^2 c^2}{64} + 4 \phi_b^2} > P > 0$. Let us now consider the expression \eqref{eq:CFTonshellOrder1} as a probability distribution. As $\phi_b$ is our clock field, we should understand \eqref{eq:CFTonshellOrder1} as function of $P$ and $ L$ for fixed values of $\phi_b$. We see that the expression \eqref{eq:CFTonshellOrder1} grows towards large values of $L $. The distribution is hence dominated by large universes.
Understanding the action as a function of $P$ will tell us something about the preferred age of the universe, parametrized by $P$ and $\phi_b$. As a function of $P$ the distribution takes on its largest value for the smallest allowed value of $P$, which is $0$. By glazing at the boundary condition \eqref{eq:metricboundaryconditionlbar} and at the Casimir action term \eqref{eq:cftcasimircontribution} we can understand where this comes from. A large universe can be guaranteed by both a large value for the initial size $b$ or by a long inflationary period corresponding to a large value for $t_c$. The Casimir term \eqref{eq:cftcasimircontribution} pushes for the former. 

It is  also an important question to understand if the distribution is normalizable. We postpone this analysis until section \ref{sec:interpretation} to include the effect of the prefactor. For generic semiclassical systems, the Wigner distribution is peaked at classical solutions of the equations of motion. In our case, we got a result by assuming a late time/old universe approximation, for which the classical solutions satisfy $ \phi_b=P+\mathcal{O}\left(c^2 e^{-2t_c}\right)$. The distribution \eqref{eq:CFTWignerdistribution} grows unbounded towards the regime where this approximation does not hold, so in this case we cannot draw any conclusion about the distribution being peaked or not at classical solutions. 

 As we have solved for the contour endpoints in a dynamical way, we should also check which contours correspond to our saddles. Inside the allowed range for $P$ we see that $t_E=\pi/4$ is reached for both $P= \phi_b$ or $c$ asymptotically large. For small $c$ we get small $t_E$ as expected. The -EAdS contour $t_E=\pi/2$, is reached when the argument of the arctan in \eqref{eq:o1saddles} asymptotes to infinity. It therefore just lies outside of the maximum bound for $P$. 
\section{Bra-Ket Wormholes and the Inflaton}
\label{sec:additionofinflaton}
In the previous two sections we determined a bra-ket solution with Wigner boundary conditions for both a purely gravitional configuration and also with the addition of a CFT. 
While pure gravity amounted to a nearly flat distribution, the addition of the CFT gave a non-zero on-shell action and hence a non-trivial distribution. However, as outlined in Appendix \ref{sec:app_wigner}, we expect the Wigner distribution for semi-classical systems to be peaked at the classical solutions of the equations of motion. The lack of this behaviour in expression \eqref{eq:CFTWignerdistribution} may be due to the quantum nature of the Casimir effect. In this section we will therefore consider a more classical matter component, for which we will indeed discover classicality of the distribution.
In an inflationary context it is natural to consider matter in the form of a slowly-rolling scalar field. While previously we considered the dilaton to act as ``time'' field, we can also consider distributions in which the inflaton plays this role. We add an action of the form:
\begin{equation}\label{eq:action_inflaton_general}
S_{f}= - \frac{\phi_0}{2}\int d^2 x \sqrt{-g} \partial^{\mu}f \partial_{\mu}f -\phi_0 \int d^2x \sqrt{-g}V(f)\,,
\end{equation}
so a minimally coupled scalar field with a potential $V(f)$. We stay in the JT limit in which we do not consider direct couplings to $\phi$, which would induce a change in the metric. However, we do endow the inflaton action with a factor $\phi_0$ as would appear from dimensional reduction. In standard FLRW coordinates we get
\begin{align}\label{eq:action_inflaton_minisuperspace}
    S_f=\phi_0\int d^2 x \, a \left(\frac{1}{2}\Dot{f}^2-V(f)\right)\,,
\end{align}
with the equations of motion 
\begin{align}\label{eq:inflatonequationsofmotion}
    \Ddot{f}+ \frac{\Dot{a}}{a} \Dot{f}+ V'(f)=0,
\end{align}
where $'$ corresponds to a derivative with respect to $f$. For simplicity we consider a linear potential $V(f)=\lambda f$, with small $\lambda$. The solution of \eqref{eq:inflatonequationsofmotion} is then given by
\begin{align}\label{eq:inflatonsolution}
    f(t)=c_1-\lambda \log(\sinh t) + c_2 \log(\tanh(t/2))\,.
\end{align}
As we would like to consider the inflaton as our clock field on a  bra-ket geometry, we set Dirichlet conditions on it $f_K=f_B=f_b$. This fixes both integration constants in \eqref{eq:inflatonsolution}. We will be again considering a contour ansatz of the form \eqref{eq:braketcontoursFLRW}. 

Now that we are using the inflaton to determine the position of the boundary, both dilaton and metric are our dynamical variables. It is thus natural to consider the following Wigner distribution:
\begin{equation}\label{eq:formalexpressionsecondWignerfunction}
    W(L,P,\Phi,Q|f_b)= \int d\Delta \phi \, d\Delta \ell \,\rho(\ell_K,\ell_B,\phi_K,\phi_B|f_b)\,e^{ i P \Delta \ell}\,e^{ i Q \Delta \phi}\,,
\end{equation}
where $\Phi=\frac{\phi_K+\phi_B}{2}$, $\Delta\phi=\phi_K-\phi_B$ and $Q$ is  the momentum canonically conjugate to the dilaton, defined as
\begin{equation}\label{eq:Qdefinition}
    Q= \frac{1}{4 \pi}\int d \varphi \left(\sqrt{h} K\right)_K- \frac{1}{4 \pi}\int d \varphi \left(\sqrt{h} K\right)_B\,.
\end{equation}

For the aforementioned linear potential under assumption of Dirichlet conditions the inflaton action on-shell takes on the simple form
\begin{align}\label{eq:inflaton_action_Dirichlet}
    S_f= -\phi_0 \int d^2 x \frac{\lambda}{2}\, a f + \phi_0 \pi f_b \Delta( a \Dot{f}).
\end{align}
This has to be added to the appropriate gravitational action for these boundary conditions which is \eqref{eq:actionforinflatonastime}, to give a total action
\begin{align}
     iS_{W(L,Q,\Phi,P|f_b)}=iS_{W(L ,Q,\Phi,P)}+iS_f\,.
\end{align}
Performing a late-time expansion on the contour \eqref{eq:braketcontoursFLRW} leaves us with the following inflaton action on the off-shell bra-ket background
\begin{align}\label{eq:offshellactionwithinflaton}
    i S_f=&2  \pi \phi_0\, b\, f_b \, \lambda \sin t_E e^{t_c} +\pi  b \lambda ^2 \phi _0 e^{t_c} \sin (t_E)-4 b \lambda ^2 (\pi -t_E) t_E \phi _0+\mathcal{O}\left(e^{-t_c}\right)\,.
\end{align}
We will always drop terms of order $e^{-t_c}$ in this section.  
The bulk solution for the metric remains the same as before, however, now we can impose two boundary conditions on it. On the contour \eqref{eq:braketcontoursFLRW} they are of the form
\begin{align}\label{eq:QFLPboundaryconditionsmetric}
   L&= b \cos(t_E)\sinh(t_c)\,,\nonumber\\
   Q&=  b \cos(t_E)\cosh(t_c)\,.
\end{align}
These equations can be used to eliminate two constants, for example:
\begin{align}\label{eq:gravitationalsectorFLQP}
    b&=\frac{1}{\cos t_E}\sqrt{Q^2-L^2}\,,\\
    e^{t_c}&=\frac{\sqrt{Q+L}}{\sqrt{Q-L}}\,.\label{eq:LQFPmodeltconshellvalue}
\end{align}
To stay in a real and positive range for $t_c$ we can only consider $Q > L$. Taking the appropriate gravitational action \eqref{eq:actionforinflatonastime} and the inflaton action \eqref{eq:offshellactionwithinflaton}, rewriting in terms of boundary conditions \eqref{eq:QFLPboundaryconditionsmetric}, we arrive at the following expression
\begin{align}\label{eq:WQFLPwithinflatonoffshellaction}
    i S(t_E)&= 2\left( \Phi L-P Q\right)\tan (t_E)+ \pi  \lambda  \phi _0 (L+Q) \left(2 f_b+\lambda \right)\tan (t_E)\nonumber\\
    &-4 \lambda ^2  \phi _0 \sqrt{Q^2-L^2}  (\pi -t_E) t_E \frac{1}{\cos t_E}+\mathcal{O}\left(e^{-t_c}\right)\,,
\end{align}
which is to be extremised over $t_E$. Let us first consider the $\lambda$ to zero limit. We see that $t_E$ essentially functions as a Lagrange multiplier. Extremisation results in a constraint on our phase space variables
\begin{equation}\label{eq:FLQPdeltafunction}
  \Phi L-P Q=0\,,
\end{equation}
thus the Wigner distribution appears to be proportional to a $\delta$-function. Since we would like to find an honest classical solution, let us consider corrections due to small $\lambda$, which will regulate the $\delta$-function and give us a nicer, smooth distribution.

 A natural starting point is the expansion around the pure gravity solution, $t_E=0$, for which the action takes on the form
\begin{align}\label{eq:offshellexpansioninflatonandgravity}
    i S(t_E)&= t_E  \left(2(\Phi L - P Q)+\pi  \lambda  \phi _0 (2 f_b+\lambda ) (L+Q)-4 \pi  \lambda ^2 \phi _0 \sqrt{Q^2- L^2} \right)\nonumber\\&+4 \lambda ^2 t_E^2 \phi _0 \sqrt{Q^2- L^2}\,,
\end{align}
where higher orders in $t_E$ are dropped. Let us now perform the extremisation with respect to $t_E$ such that we end up with
\begin{equation}\label{eq:onshellvaluetE}
    t_E=\frac{2(P Q-\Phi L)-\pi  \lambda  \phi _0 (2 f_b+\lambda ) (L+Q)}{8 \lambda ^2 \phi _0 \sqrt{Q^2-L^2}}\,,
\end{equation}
and
\begin{align}\label{eq:qflpaction}
    iS=-\frac{\left(2(\Phi L-P Q)+\pi  \lambda  \phi _0 (L+Q) \left(2 f_b+\lambda \right)\right){}^2}{16 \lambda ^2 \phi _0 \sqrt{Q^2-L^2}}.
\end{align}
Hence we get a Gaussian distribution in the combination of phase space variables corresponding to $t_E$:
\begin{equation}\label{eq:LFQPdistribution}
W(L,Q,\Phi,P|f_b)= \mathcal{N}_{L,Q,\Phi,P}\,\text{exp}\left(-\frac{\left(2(\Phi L-P Q)+\pi  \lambda  \phi _0 (L+Q) \left(2 f_b+\lambda \right)\right){}^2}{16 \lambda ^2 \phi _0 \sqrt{Q^2-L^2}}\right)\,,
\end{equation}
where again we kept an undetermined normalization factor. 

The distribution attains its maximum on a three-dimensional hypersurface in phase space:
\begin{align}\label{eq:WFLPQ_max_hypersurface}
    \Phi L-P Q + \frac{\pi  \lambda  \phi _0}{2} (L+Q) \left(2 f_b+\lambda \right)=0\,.
\end{align}
On this surface, $t_E=0$. To compare to what we expect from the generic semiclassical approximation for Wigner distributions, it is instructive to look at the classical solution for the dilaton. The backreacted equations for the dilaton are of the form 
\begin{align}
    \ddot{\phi}-\phi + 2\pi \phi_0 \left(\frac{1}{2}\dot{f}^2-\lambda f\right)&=0\,,\\
    \phi-\frac{\dot{a}\dot{\phi}}{a} \left(\frac{1}{2}\dot{f}^2 + \lambda f\right)&=0\,,
\end{align}
with the inflaton given by \eqref{eq:inflatonsolution}. This results in a lengthy expression for the dilaton, which we do not show here in detail.
In the $t_E$ to zero limit and at large $t_c$, this leaves us with the following boundary expressions for the two dilaton phase space variables
\begin{align}
    \Phi&= A \cosh(t_c) + \frac{\lambda}{4}\left(-2 f_b \pi \phi_0 + \pi \lambda \phi_0\right)\,,\\
 P&= A \sinh(t_c) + \lambda^2 2 \pi \phi_0\,.
\end{align}
We see that the relation \eqref{eq:WFLPQ_max_hypersurface} is satisfied by these values. Hence, we find the Wigner distribution is peaked in a region of phase space close to the classical solution. As we discuss in Appendix \ref{sec:app_wigner}, this is what one expects from generic semiclassical systems, while opening of the contour, parametrized by $t_E$, corresponds to quantum effects. In the remaining three directions our distribution is flat in this approximation. These directions parametrize different classical solutions. We can choose them as the initial values of the metric and the dilaton, as well as the age of our universe. In particular, we see that the distribution is flat with respect to the age, as in the case of no matter.
We can also consider the addition of the CFT on top of the inflaton. We keep the inflaton and CFT decoupled, but both backreacting on the dilaton. The action \eqref{eq:offshellexpansioninflatonandgravity} is supplemented by the contribution of the effective CFT action \eqref{eq:cftcasimircontribution}, where now $b$ can be substituted using \eqref{eq:gravitationalsectorFLQP}.
With the new gravitational boundary conditions \eqref{eq:QFLPboundaryconditionsmetric} the CFT takes the same form as above \eqref{eq:cftcasimircontribution}.
Extremising with respect to $t_E$ leaves us with 
\begin{align} \label{eq:teinflaton+CFT}
    t_E^*=-\frac{8 \left(\Phi L -P Q + 4\pi  \lambda  \phi _0 (L+Q) \left(2 f_b+\lambda \right)\right)}{\sqrt{Q^2-L^2} \left(\pi  c+32 \lambda ^2 \phi _0\right)}\,,
\end{align}
and the Wigner distribution:\footnote{The prefactor can be slightly different in the presence of CFT, but we still use the same symbol.}
\begin{align}\label{eq:LQFPwithcft}
    &W_{\text{CFT}}(L,Q,\Phi,P|f_b)\nonumber\\
&=\mathcal{N}_{L,Q,\Phi,P}\exp\left(\frac{1}{4} \pi  c \sqrt{Q^2-L^2}-\frac{2 \left(2(\Phi L-P Q)+\pi  \lambda  \phi _0 (L+Q) \left(2 f_b+\lambda \right)\right){}^2}{\sqrt{Q^2-L^2} \left(\pi  c+32 \lambda ^2 \phi _0\right)}\right).
\end{align}
Let us consider the full four-dimensional gradient. We should distinguish between the behaviour for $\Phi, P$ and $Q, L$. Clearly the extremisation with respect to either $\Phi$ or $P$ still results in \eqref{eq:WFLPQ_max_hypersurface}, however, we now have a factor coming from the CFT action that weighs solutions differently  with different values of $\sqrt{Q^2-L^2}$. Similarly to the distribution discussed in section \ref{sec:braketwhandconformalmatter}, it pushes for large values of the initial size of the universe.
We should also be aware of the different scalings of the two terms in \eqref{eq:LQFPwithcft}. As the Casimir term constitutes a loop effect, it can be naturally made smaller than the classical inflaton contribution. In Appendix \ref{app:units} we restore units of Planck mass $M_p$ and de Sitter radius $R_{dS}$ starting from the higher dimensional geometry $dS_2 \times S^2$.
The result is
\begin{align}\label{eq:wignerFLPQwithCFTandunits}
    &W_{\text{CFT}}(L,Q,\Phi,P|f_b)=\mathcal{N}_{L,Q,\Phi,P}\nonumber\\
&\times\exp\left(\frac{1}{4} \pi  c \sqrt{Q^2-L^2}-\frac{2  \left[2(\Phi L-P Q)+\pi  \lambda \, R_{dS}^4 (L+Q) \left(2 f_b+\lambda \, R_{dS}^2 \right)\right]^2}{\sqrt{Q^2-L^2} \left(\pi  c+32 \lambda^2 \, R_{dS}^6 \right)}\right)\,.
\end{align}
Where the dilaton $\Phi$ and derivative $P$ get a scaling of order $M_p^2 R_{dS}^2$. We clearly see that the purely gravitational contribution is enhanced by $M_p^2 R^2_{dS}$ with respect to the CFT contribution. The inflaton term is enhanced via $f_b$. In general we can think of this boundary value of the inflaton as being order one in Planck units, as typical for inflationary potentials, making the full gravity plus inflaton contribution dominate over the one-loop CFT contribution (unless $c$ itself scales appropriately with $M_p$). 

\section{One-loop Determinant}\label{sec:oneloopdeterminant}
So far we have focused on determining the action of (connected) saddlepoint contributions to different Wigner functions both with and without matter. Naturally we expect corrections as expansions around these saddles. Such one-loop contributions may play an important role in the gravitational path integral. The most pertinent example is the double-cone wormhole of \cite{Saad:2018bqo,Saad:2019lba}. As the leading connected contribution to the spectral form factor $\overline{|Z(\beta + i T)|^2}$ it exhibits late time ``ramp'' behaviour typical of a chaotic system. Moreover, it demonstrates that topologically suppressed contributions may dominate over disconnected connected geometries via loop effects in some regimes. In previous sections we established a more geometric approach in determining various functions, i.e we gave an explicit parametrization of the contour with the appropriate Wigner boundary conditions. In this section we take a different path which is more suited to determining all loop effects. We start by fixing an off-shell density matrix, which is a function of both its boundary conditions and of the cycle $b$. To arrive at a connected contribution to the Wigner function we then perform both the integration with respect to the appropriate boundary variables and the internal variable $b$. We expect changes to the prefactor to come from three different sources. Firstly, there are corrections coming from the Schwarzian mode which already have to be included in the off-shell density matrix. Secondly, there are fluctuations around the joint saddlepoint of the internal $b$ integration and the appropriate Wigner transform. Thirdly, for connected geometries we have to integrate over the internal length parameter $b$ and the relative twist $\tau$ between the glued surfaces. As the surface is invariant under a twisting by proper distance $b$, we consider the appropriate measure to be $\int_{0}^{\infty}b \, db $ \cite{Saad:2019lba}. 
Let us start by determining the prefactor of the simplest setup: the vacuum Wigner function $W( L,P;\phi_b)$. The first step is determining the Schwarzian corrected off-shell density matrix. On both boundaries we set the boundary conditions on the induced metric:
\begin{align}\label{eq:Schwarzianbcmilne}
b \sinh(t_K(u))\varphi_K' &= \ell_K\,,\\
b \sinh(t_B(u))\varphi_B' &= \ell_B\nonumber\,,
\end{align}
where $'$ denotes a derivative with respect to the intrinsic boundary coordinate $u$.
Performing the path integral over the Schwarzian modes, which is done in detail in Appendix \ref{app:Schwarziancorrections}, results in the expression 
\begin{align}\label{eq:densitymatrix1}
\rho^{(b)}(\ell_K,\ell_B|\phi_b)&= \frac{1}{2}\sqrt{\frac{\phi_b^2}{\ell_K \ell_B}}\text{exp}\left(i \phi_b(\ell_B-\ell_K) + i b^2 \frac{\phi_b}{2}\left( \frac{1}{\ell_B} - \frac{1}{\ell_K}\right)\right)\,,\nonumber\\
&= \sqrt{\frac{ \phi_b^2}{4 L^2-\Delta \ell^2}}\text{exp}\left(- i \phi_b \Delta \ell + i b^2 \phi_b\left(\frac{2\Delta \ell}{4 L^2-\Delta \ell^2}\right)\right)\,,
\end{align}
where we have inserted Wigner variables and expanded at late times. To be clear this is an off-shell expression as the $b$ mode is still unfixed, which we indicate by a superscript. This result is one-loop exact, as can be shown from localization \cite{Stanford:2017thb} or gravity arguments \cite{Anninos:2021ydw}. Now in order to calculate the on-shell Wigner function we both have to perform the Wigner transform and the internal integration with the measure mentioned above, such that we end up with the double integral
\begin{align}\label{eq:WignerPLviadensity}
W(P, L|\phi_b)&=\int_{- 2 L}^{ 2 L}(d\Delta \ell) \int_0^{\infty} b db \,\rho^{(b)}(\ell_K,\ell_B|\phi_b)e^{ i\Delta \ell P  }\,.
\end{align}
We have restricted the range of integration such that both $\ell_K$ and $\ell_B$ remain positive by assumption.\footnote{Of course as is the case when including matter this still allows for contours which may include complex saddlepoints.}
We perform the double integration via saddlepoint. The only non-singular solution to the equations is:
\begin{align}\label{eq:saddleforPLbphib}
\Delta \ell &= 0\,,\\
b&= \frac{ L\sqrt{ 2(\phi_b-P)}}{\sqrt{\phi_b}}\equiv b^*\nonumber\,.
\end{align}
These are the same saddlepoint values as in \eqref{eq:onshellvalues vacuum}, such that we end up with the same on-shell action, which for this configuration is zero. Now having included both Schwarzian fluctuations, and a measure on the space of solutions, which are both to be evaluated on the above saddlepoint value, we must also include the first Gaussian fluctuations around this saddle. Only the mixed derivative is non-zero on the saddle \eqref{eq:saddleforPLbphib} and is purely imaginary. Therefore convergence of the integral requires deformation of the contour of integration. We assume that the correct integration contour in our gravitational path integral is such that this procedure is correct. Let us write the result as a product of the aforementioned three contributions: the measure on phase space on the saddlepoint value, the Schwarzian prefactor and the Gaussian fluctuations
\begin{equation}
W(L,P|\phi_b) = b^{\ast}\frac{\phi_b}{2 L} \frac{2 \pi L }{ \sqrt{\phi_b}\sqrt{2(\phi_b-P)}}  \,.  
\end{equation} Overall we end up with 
\begin{align}\label{eq:WignerPLbarvacuumwithprefactor}
W(L,P|\phi_b) &=  \pi L\,.
\end{align}
Hence we have determined the prefactor $\mathcal{N}_{ L,P}$ of the vacuum Wigner function \eqref{eq:vacuumLPphibWignerfuncton}.
Somewhat analogous to the ``ramp'' behaviour of \cite{Saad:2018bqo,Saad:2019lba}, we see a simple linear growth. One can interpret this as a (non-normalizable) distribution on phase space that prefers large universes.\footnote{To properly discuss normalizability of the distribution we need to specify the integration measure over $L$. It was pointed out in \cite{Cotler:2024xzz}, and in the appendix G of \cite{Maldacena:2019cbz} that with our definition of the wave function the measure should have an additional factor of $1/L$, or of $\phi_b/L$. This is also analogous to putting such factor in the Wigner distribution itself. This factor is not going to play a role in what follows since it does not change normalizability properties of the distribution, and it still prefers large universes. Moreover this factor should be common to the connected and disconnected contributions, so it would not change their relative weights.}  If we add conformal matter, we expect 
 the saddlepoint value of $\Delta \ell$ to only change by a small amount. There is a subtlety because only the zero momentum CFT states contribute to the partition function due to the integral over the twist angle. Still the dominant contribution to the prefactor comes from the determinant of the fluctuations which is not affected by this. Hence, we expect that the expression \eqref{eq:CFTWignerdistribution} in that regime also has a linear in $L$ prefactor\footnote{There is an additional contribution to the prefactor which comes form the fact that the momentum constraint projects the matter contribution to the zero-momentum sector. At large $c$ this factor can be evaluated via a saddle point integral over the gravitational mode that corresponds to the relative shift of coordinates on bra and ket boundaries. It turns out to be subleading as compared to the exponent and we will not keep track of it.} 
\begin{equation}\label{eq:upgradedresultPLCFT}
    W_\text{CFT}(L,P|\phi_b)\approx\pi L \,\text{exp}\left(\frac{\pi  c L  \sqrt{4 (\phi_b^2-P^2)+\frac{\pi ^2 c^2}{16}}}{ \left( \phi _b+P\right)}\right)\,.
\end{equation}
Our final goal is to determine the prefactor of the Wigner function $W(L,Q,\Phi,P|f_b)$. 
While a direct calculation would involve including the one-loop determinant of the inflaton and then performing both the Wigner transform and the internal integrations, we consider a more indirect approach.\footnote{See references \cite{Saad:2019pqd,Mertens:2017mtv,Jafferis:2022wez} for JT gravity with matter.} The basic idea is that we can match inverse Wigner transforms of our result for $W(L,Q,\Phi,P|f_b)$, formula \eqref{eq:LFQPdistribution}, to $W(L,P|\phi_b)$ as defined via the density matrix \eqref{eq:WignerPLviadensity}, in the diagonal and $\lambda$ to zero limit. 
This constrains the prefactor $\mathcal{N}_{L,Q,\Phi,P}$. Let us be more explicit. We define the following inverse transform \begin{align}\label{eq:LPphifbwignerinverse}
    W(L,P;\phi_K,\phi_B|f_b)=\int_{ L}^{\infty}dQ \, W( L,Q,\Phi,P|f_b) e^{- i Q \Delta \phi}\,.
\end{align}
Note that here we have restricted the inverse transform to the allowed region for the variable $Q$ we determined in section \ref{sec:additionofinflaton}. This object is a density matrix in the dilaton boundary values. We expect the following formula to hold
\begin{equation}
    W(L,P;\phi_b,\phi_b|f_b)|_{\lambda \rightarrow 0} = W(L,P|\phi_b)\,,
\end{equation}
with $W(L,P|\phi_b)$ as given in \eqref{eq:WignerPLbarvacuumwithprefactor}. For large phase space variables the left-hand side is calculable in a saddlepoint approximation. As a $\lambda$-expansion the saddlepoint is at
\begin{equation}\label{eq:Qsaddlepointforinversetransform}
Q^{\ast}= \frac{\Phi  L}{P}+\mathcal{O}(\lambda)\,.
\end{equation}
In the diagonal limit the action vanishes such that the result will merely be a prefactor. This prefactor consists of the as of yet undetermined normalization factor of the Wigner function $W(L,P;\phi_b,\phi_b|f_b)$ on the saddlepoint value \eqref{eq:Qsaddlepointforinversetransform} and Gaussian fluctuations of $Q$ around the saddlepoint value of the action. Hence,
\begin{align}
&W(L,P;\phi_b,\phi_b|f_b)\nonumber\\=&\mathcal{N}_{ L,Q,\Phi,P}(Q^{\ast})\frac{ 2 \sqrt{\pi}\lambda \sqrt{\phi_0}\sqrt{ L}(\phi_b^2-P^2)^{1/4}}{P^{3/2}}\,.
\end{align}
Here we have identified the Wigner variable $\Phi$ with the diagonal density matrix variable $\phi_b$.
This must now match the expression \eqref{eq:WignerPLbarvacuumwithprefactor} leading to the constraint
\begin{align}\label{eq:constraintsonqflpprefactor1}
\mathcal{N}_{Q,\Phi, L,P}(Q^{\ast})&= \frac{\sqrt{\pi}P^{3/2}\sqrt{ L}}{2\lambda \sqrt{\phi_0}(\phi_b^2-P^2)^{1/4} }\,.
\end{align}
 In the vicinity of the distribution maximum we can thus use the following expression for the prefactor:
\begin{equation}\label{eq:answerqflpprefactor}
    \mathcal{N}_{Q,\Phi,L,P}=\frac{\sqrt{\pi}P  L}{2\lambda\sqrt{\phi_0}(Q^2- L^2)^{1/4}}\,.
\end{equation}
As a more detailed check of this procedure we extend the matching of the two approaches to off-diagonal matrix elements in Appendix \ref{app:inverseWigner}. 

There is also another type of Wigner distribution that we could consider in order to calculate the prefactor. It is the one where we fix the sum of the dilaton at the two boundaries $\Phi=\frac{\phi_K+\phi_B}{2}$ and we do the Fourier transform of the density matrix with respect to the difference of dilaton values. In this case we select the boundary length as a clock, i.e. set it equal to a common value $\ell_B=\ell_K=\ell_b$ at the two boundaries. The formal expression for this is
\begin{equation}\label{eq:formalexpressiontemperatureWignerfunction}
    W(\Phi,Q|\ell_b)=\int_{-2 \Phi}^{2 \Phi} d \Delta \phi \rho(\phi_K,\phi_B|\ell_b)e^{i Q \Delta \phi }\,.
\end{equation}
Also this type of Wigner distribution admits a path integral representation for which the action is given in Appendix \ref{secapp:gravitationalaction}. We can use similar logic as above to determine the prefactor of such a Wigner distribution without matter. The object of study is the following integral
\begin{equation}\label{eq:QF}
W(\Phi,Q|\ell_b)=\int_{-2 \Phi}^{2 \Phi} d \Delta \phi  \int_{0}^{\infty}b db \rho^{(b)}(\phi_K,\phi_B|\ell_b) e^{ i\Delta \phi Q} \,,
\end{equation}
where again we have restricted the range of the Wigner integration to keep to positive values for $\phi_K$ and $\phi_B$. Here we have introduced a new type of density matrix $\rho^{(b)}(\phi_K,\phi_B|\ell_b)$, which is off-shell due to the unfixed internal mode. While it is a matrix in the two dilaton values, the boundary length is set the same on both bra and ket (thus we use $\ell_b$ as time). Performing the Schwarzian path integral gives
\begin{equation}
\rho^{(b)}(\phi_K,\phi_B|\ell_b)= \frac{\sqrt{4\Phi^2-\Delta \phi^2}}{4 \ell_b}\text{exp}\left(-i \ell_b\Delta \phi - \frac{i b^2 \Delta \phi}{2 \ell_b} \right)\,.
\end{equation}
The saddlepoint is then found at 
\begin{align}
b^{\ast}&= \sqrt{2\ell_b(Q-\ell_b)}\,,\\
\Delta \phi &=0\nonumber\,,
\end{align}
which agrees with the ``direct'' approach.\footnote{By which we mean an explicit contour parametrization as performed in previous sections. Let us however note that off-diagonal matrix elements in principal include real displacements between bra and ket.} Again assuming an appropriate contour rotation in order to make the Gaussian integral well-defined, we arrive at
\begin{equation}
W(\Phi,Q|\ell_b) = b^{\ast}\frac{\Phi}{2\ell_b} \frac{2 \pi \ell_b}{\sqrt{2\ell_b(Q- \ell_b)}}\,,
\end{equation}
where $b^{\ast}$ now refers to the above saddlepoint value. Overall we then get
\begin{align}\label{eq:WignerQFLvacuumwithprefactor}
W(\Phi,Q|\ell_b)&=\pi \Phi \,.
\end{align}
Here too we expect that this object can be matched to an inverse transform of $W(L,Q,\Phi,P|f_b)$. That is, it should hold that
\begin{equation}
W(Q,\Phi;\ell_b,\ell_b|f_b)|_{\lambda \rightarrow 0} = W(\Phi,Q|\ell_b)\,.
\end{equation}
Let us therefore consider 
\begin{equation}\label{eq:QFLbwignerinverse}
    W(Q,\Phi;\ell_K,\ell_B|f_b)= \int_{0}^{\Phi} dP  \,W( L,Q,\Phi,P|f_b)e^{-i P \Delta \ell} \,,
\end{equation}
This is a density matrix in the length variables $\ell_K$ and $\ell_B$. We have restricted to the allowed region for the integration varaible. This integral exhibits a saddlepoint at 
\begin{equation}\label{eq:saddlepointvaluePforinversetrafo}
    P^{\ast}=\frac{\Phi L}{Q} + \mathcal{O}(\lambda)\,.
\end{equation}
 The result of \eqref{eq:QFLbwignerinverse} should match onto the previous result \eqref{eq:WignerQFLvacuumwithprefactor} when taking the diagonal limit and sending $\lambda$ to zero. The action vanishes on the saddle. We display the prefactor as a product of the normalization of $W(L,Q,\Phi,P|f_b)$ on the saddlepoint \eqref{eq:saddlepointvaluePforinversetrafo} and the determinant of Gaussian fluctuations of $P$
 \begin{equation}
    W(Q,\Phi;\ell_b,\ell_b|f_b)=\mathcal{N}_{Q,\Phi, L,P}(P^{\ast})\frac{2 \sqrt{\pi}\lambda\sqrt{\phi_0}(Q^2-\ell_b^2)^{1/4}}{Q}\,,
\end{equation}
where we have identified $L$ with $\ell_b$.
Matching to \eqref{eq:WignerQFLvacuumwithprefactor} supplies the constraint
\begin{align}\label{eq:constraintsonqflpprefactor2}
\mathcal{N}_{Q,\Phi, L,P}(P^{\ast})&= \frac{\sqrt{\pi}\Phi Q}{2\lambda \sqrt{\phi_0}(Q^2-\ell_b^2)^{1/4}}\,.
\end{align}
At leading order in late times this condition is satisfied by \eqref{eq:answerqflpprefactor} providing us with a cross-check of this expression. We can therefore upgrade the result for $W(L,Q,\Phi,P|f_b)$, formula \eqref{eq:LFQPdistribution}, to 
\begin{align}\label{eq:LFQPdistributionupgraded}
&W( L,Q,\Phi,P|f_b)\\
=& \frac{\sqrt{\pi}P  L}{2\lambda\sqrt{\phi_0}(Q^2- L^2)^{1/4}}\text{exp}\left(-\frac{\left(2(\Phi L-P Q)+\pi  \lambda  \phi _0 (L+Q) \left(2 f_b+\lambda \right)\right){}^2}{16 \lambda ^2 \phi _0 \sqrt{Q^2-L^2}}\right)\nonumber\,.
\end{align}
This holds at late times and small $\lambda$ and only near the extremum of the gaussian. In the next section we will discuss where this distribution is peaked and what its probabilistic interpretation is.
\section{Probabilistic Interpretation and Observables}\label{sec:interpretation}
\subsection{Experience of an ``inflationary'' observer}
\label{observer}
We now come to a more thorough interpretation of the results described so far. The results established here are to be understood as distributions for the gravitational zero modes of the early universe, that is the state of the universe after the inflationary phase in our toy model. We can now imagine a local observer that does measurements in some part of this inflationary patch, in analogy with us measuring CMB fluctuations. What kind of predictions can such an observer make based on the distribution \eqref{eq:LFQPdistributionupgraded}? As noted above, one prediction is that at late times the universe was evolving close to classical equations of motion. To understand the meaning of the distribution on the entire phase space better it is convenient to introduce the following ratios:
\begin{equation} \label{eq:uvcoordinates}
   U\equiv\frac{P}{\Phi}\,, \quad V \equiv \frac{Q}{L}\,,
\end{equation} 
that do not depend on the overall size of the universe and of the sphere (dilaton). Instead, their values are directly related to how long the universe evolved following the Euclidean regime (age of the universe). In these variables the distribution reads
\begin{align}\label{eq:wignersmallpq}
   &W(L,\Phi,U,V|f_b)\nonumber\\=& 2\sqrt{ \pi} \frac{\kappa}{(V^2-1)^{1/4}}\exp\left\{ - \kappa^2 V^2\, \frac{ \left[2(\frac{1}{V}-U) +\pi \lambda  \frac{\Phi}{\phi_0} \, \left(\frac{1}{V}+1\right)(2 f_b+\lambda)\right]^2}{(V^2-1)^{1/2}}\right\} \,,
\end{align}
where 
\begin{equation}
    \kappa\equiv \frac{\Phi \sqrt{L}}{\sqrt{16 \lambda^2 \phi_0}}\,.
\end{equation}

Now imagine that $L$ and $\Phi$ are both large, then local observables cannot depend on them, while they can still depend on $U$ and $V$ because these are locally measurable properties of the background.\footnote{To justify that local observables do not depend on $\Phi$ we assume that it has the meaning of a sphere size, and that observables are also local on the sphere.} Consider some local observable $\hat {\mathcal{O}}$. In order to calculate its expectation value one needs to take its Wigner transform and integrate over the phase space with the Wigner distribution. Generically this observable will depend on $U$ and $V$ similarly to how CMB fluctuations depend on the evolution of the scale factor of our universe, as well as on some local modes that we collectively denote as $X_i$.\footnote{It does not matter at this point whether to use Wigner formalism, or usual formalism for short modes, we use Wigner in order to make notation more homogeneous.\label{ref:footnote}} Then

\begin{equation}
 \langle\hat {\mathcal{O}}\rangle=\int d\Phi\, {\Phi^2} \int dL\,   \int_{0}^{1} dU\, \int_{1}^{\infty} dV\,\int dX_i\,W(L,\Phi,U,V,X_i|f_b) \mathcal{O}_W(U,V,X_i)\,,
\end{equation}
Where we included additional factor of $\Phi/L$ in the measure, according to the discussion in \cite{Cotler:2024xzz}, \cite{Maldacena:2019cbz}.

Let us first take an integral over $U$. In the small $\lambda$ and large $\Phi$, $L$ limit it is a very narrow Gaussian, basically a $\delta$-function of the classical solution:
\begin{equation}
\label{deltadistr}
W\propto\frac{1}{V}\delta(U-U_{\text{cl}}(V))\,.
\end{equation}

We thus get

\begin{equation}
 \langle\hat {\mathcal{O}}\rangle=\int d\Phi\, \Phi^2 \int dL\,     \int_{1}^{\infty} dV\,\int dX_i\,\frac{1}{V}\tilde{  W} (U_{\text{cl}},V,X_i|f_b )\mathcal{O}_W(U_{\text{cl}},V,X_i)\,,
\end{equation}
where $\tilde{W} (U_{\text{cl}},V,X_i|f_b )$ is the distribution of local modes on the classical solution of $U$ and $V$ that we chose to parametrize by $V$. Note that the dependence on $L$ and $\Phi$ got factorised. Corresponding integrals are also IR divergent, however, this divergence is the same for all local observables and we can thus ignore it. One more comment is in order: even though we extended the integral over $V$ all the way to infinity, we have control over the prefactor in our distribution only for old universes, corresponding to $V$ close to 1, since we have on our classical solutions $t_c\propto -\frac{1}{2}\log(V-1)$. Thus really the expression we have is 
\begin{equation}\label{eq:expvalWignerUV}
 \langle\hat {\mathcal{O}}\rangle\propto \int_{1}^{1+\epsilon} dV\,\int dX_i\,\tilde{  W} (U_{\text{cl}},V,X_i|f_b )\mathcal{O}_W(U_{\text{cl}},V,X_i)\, + \text{young universes contribution}\,.
\end{equation}
Let us introduce the notation $\langle\hat {\mathcal{O}}\rangle_V$ for an expectation value of some operator given a fixed value of $V$, then
\begin{equation}
\label{localVint}
 \langle\hat {\mathcal{O}}\rangle\propto \int_{1}^{1+\epsilon} dV\langle\hat {\mathcal{O}}\rangle_V\, + \text{young universes contribution}.
\end{equation}

This result means that local observables should be calculated on classical solutions for the background modes and then integrated over the phase space variable $V$ which corresponds to the Hubble parameter measured at $f=f_b$. This integration has the meaning of averaging over the length of Lorentzian evolution (age of the universe). For old universes this measure is approximately flat with respect to this variable $V$.\footnote{If we include into consideration the fact that the inflaton potential has a finite range, there will be some maximal number of e-foldings allowed, effectively introducing a lower bound on integration over $V$ which is larger than 1.\label{ref:inflationend}} If we include CFT matter, as in \eqref{eq:LQFPwithcft}, the probability of young universes will be enhanced, however, our calculation of the prefactor is not valid in this regime. It would be interesting to use the methods of \cite{Iliesiu:2020zld,Stanford:2020qhm} to compute the distribution in the regime when the boundary is not at late times. If we look at the distribution as a number of e-foldings $N_e\sim t_c$ we will get the measure of the form $e^{-2 N_e}dN_e$. Thus it has a similar issue as the 4D HH wave function, which does not correctly predict the length of inflation, however, the preference for young universes is much milder and the technical reason is different. In the HH case, the preference for short inflation is already at the level of the action. Here, as generic for Wigner distributions (see Appendix \ref{sec:app_wigner}), the classical action vanishes on all classical solutions and we need to look at the prefactor which is not as universal and depends on quantum corrections. It could be that in some models these corrections lead to a distribution peaked around old universes. 

 If we project on the region of phase space corresponding to old enough universes, our distribution has a meaningful probabilistic interpretation. Below we will show an example computation of a local observable, namely correlation functions of a free scalar field on our wormhole background. In section \ref{sec:temperatureastime} we will give an example of how one can motivate projecting onto old universes in our setup.

Until now we considered compact universes, even though we saw that the overall size $L$ tends to be large. Classically we can consider geometries that correspond to a non-compact universe and take $L\to \infty$. In this limit the value of $a$ is not meaningful, however, $\dot a/ a$ is still observable. It is somewhat puzzling how to think about this from the phase space point of view, because the phase space is naively odd-dimensional. Nevertheless, it is reasonable to assume that the delta-functional Wigner distribution \eqref{deltadistr} that depends on $U$ and $V$ only is still the right answer in this situation, assuming that $\Phi$ is also large.

Above we found the classical solution with a connected geometry corresponding to Wigner boundary conditions, while there was no solution with boundary conditions corresponding to a density matrix. If our distribution were an integrable function on phase space with a smooth Fourier transform the classical saddle for the density matrix would also have existed. As we just discussed, while we have a reasonable probabilistic interpretation for our distribution, it is not normalizable in two directions corresponding to large volume and large dilaton, and in addition to this exhibits an unknown behaviour in the region of phase space corresponding to young universes. Thus our phase space approach suggests the regularization which is needed in order to make the density matrix also well defined: we need to regulate the IR divergences corresponding to large $L$ and $\Phi$, as well as the UV divergence corresponding to early times. We can imagine that both are regulated by some non-perturbative effects, possibly of a different nature. For now we proceed with studying our distribution in the region of phase space where our classical solutions are under control, and we provide some details on inverse Wigner transforms in Appendix~\ref{app:inverseWigner}.

\subsection{Momentum modes and power spectrum}
Let us now consider perturbations of the inflaton field on our bra-ket wormhole background. They can be thought of as an analogue of CMB anisotropies in our toy model of inflation. We treat perturbations as a free scalar field with action
\begin{equation}\label{Sf}
S_{f}=-\int d^2 x\sqrt{g}(\partial_{\mu} f) (\partial^{\mu} f)\,.
\end{equation}
 Let us consider a contribution of a single momentum mode 
\begin{equation}
f(\eta,\varphi)=f_k(\eta)e^{i k \varphi}+f_{-k}(\eta)e^{-i k \varphi} \,,
\end{equation}
where $k$ is the comoving momentum. We assume $k\gg1/b$ so that we can treat it as a continous variable. The distribution for the relevant background modes is given in  \eqref{eq:wignersmallpq}. As we discussed, it is very narrowly peaked around the classical solutions which correspond to a closed contour without any Euclidean displacements of the endpoints. We will work in this approximation. In fact, in conformal coordinates the metric drops out from the action \eqref{Sf} and the solutions of equation of motion for $f_k(\eta)$ are the same as on a flat torus. If we fix the boundary values of the mode to be $f_{k,b}$ the on-shell action takes the form           
\begin{equation}
 i S_{f,k}= -  2 \pi f_{k,b} f_{-k,b} k \tanh{\frac{k \pi}{2}}\,,
\end{equation}
and the correlation function is then given by
\begin{align}
    \langle \hat f_k \hat f_{-k} \rangle_V= \frac{1}{4 \pi k \tanh{\frac{k \pi}{2}}}\,.
\end{align}
The corresponding physical observable is the power-spectrum as a function of physical momenta $q= - k \sinh \eta_c\approx - k \eta_c$ at late times. On our solution $\eta_c$ is related to the phase space variable $V$ as $\eta_c=-\sqrt{V-1}$ for $V$ close to 1. Making the change of variables the power spectrum reads

\begin{align}\label{eq:scalartwopointphysical}
    \langle \hat f_q \hat f_{-q} \rangle_V d q= \frac{1}{4 \pi} \coth\left( \frac{\pi q}{ 2 \sqrt{V-1}}\right)\,\frac{d q }{q}.
\end{align}
Let us keep $V$ fixed for now.
We see that for modes that have physical momenta such that $q/\eta_c \gg 1$, i.e. modes that crossed the horizon at some point during the Lorentzian evolution of our universe, we get the scale-invariant power-spectrum, the analogue of $d^3q\frac{1}{q^3}$ in four dimensions.
 For modes that still had long wavelength when our universe was in the Euclidean regime, that is with  $q/\eta_c \ll 1$ we instead get a different correlation function because they are sensitive to the overall torus topology of the bra-ket spacetime:
\begin{align}\label{eq:scalartwopointlong}
    \langle \hat f_q \hat f_{-q} \rangle_V= \frac{1}{2 \pi^2} \frac{\sqrt{V-1}}{q^2}\,.
\end{align}
This can be thought of as  a thermal spectrum at temperature $\frac{\eta_c}{\pi}$.\footnote{Correlation functions in 2d CFTs at finite temperature behave as $\langle\hat O(x) \hat O(0) \rangle\approx(\sinh{\pi T x})^{-2\Delta}$. Massless scalar is not a well-defined operator, but we can approximate it with an operator of dimension $\Delta\to0$, so that the two-point function at large distances behaves as $T x$, which leads to the Fourier transform behaving like $T /q^2$ for small $q$.} Following our discussion above, we in principle need to integrate our power spectrum over $V$, as in equation \eqref{localVint}. From the point of view of our power spectrum it will look like integrating over the scale at which the transition between the scale-invariant and thermal spectra occurs. In any given realization of the universe once this scale is measured it remains the same for all other observables because $V$ is a very classical variable. Such a measurement will implement a projection on a given value of $V$.
\subsection{Classicality}\label{subsec:interpretation_classicality}
We have in this section so far worked under the assumption that the contour opening as parametrized by $t_E$ in the contour ansatz \eqref{eq:braketcontoursFLRW} was small. At various points of the text we have emphasised that we think of this parameter as a measure of classicality, that is how close the state is to a classical state. We will now check under which condition the assumption of small $t_E$, which is required to arrive at an action of the form \eqref{eq:wignersmallpq}, holds. First, the Taylor expansion of the action is valid for small $t_E<<1$, which as can be seen from the on-shell value \eqref{eq:onshellvaluetE} is true for 
\begin{equation}\label{eq:tetaylorexpansioncondition1}
   {8 \lambda ^2 \phi _0 \sqrt{V^2-1^2}}>> |2\Phi L( U V -1) - \lambda  \phi _0 (2 f_b+\lambda ) (1+V))|\,.
\end{equation}
Having determined the distribution \eqref{eq:wignersmallpq} we should also ask if the approximation we made is valid for typical values of the parameter $t_E$ inside the spread of the distribution, that is for phase space values for which the distribution becomes $\mathcal{O}(1)$. We can use the result \eqref{eq:wignersmallpq} to determine the expectation value of $\langle\hat{t}_E^2\rangle$.\footnote{Due to the nearly Gaussian nature of the distribution \eqref{eq:wignersmallpq}, we have $\langle\hat{t}_E\rangle \approx 0$.}
Integrating the square of the expression for $t_E$, equation \eqref{eq:onshellvaluetE}, and normalizing with respect to the overall distribution results in
\begin{equation}\label{eq:expectationvaluetE^2}
 \langle\hat{t}_E^2\rangle_{V}   \propto \frac{1}{ L_{\Lambda} \lambda^2 \phi_0\sqrt{V^2-1}}\,.
 \end{equation}
Here we see that the cutoff on the $L$ integration, $L_{\Lambda}$ appears explicitly. This is because $t_E$ is not a local operator but a metric parameter.
Validity of the approximation inside the Gaussian in addition to the constraint \eqref{eq:tetaylorexpansioncondition1} also leads to the demand that the above expectation value is parametrically small, which is achieved by
\begin{equation}\label{eq:tetaylorexpansioncondition2}
  L_{\Lambda} \lambda^2 \phi_0\sqrt{V^2-1}>>1\,.   
\end{equation}
For a given $L_{\Lambda}$ this puts a lower constraint on $V$. Such a lower bound can be easily implemented by modifying the inflation potential,  see footnote \ref{ref:inflationend}. For the solutions we are considering this implies that geometrically the parameter $b$, which we can think of roughly as the size of the universe in the Euclidean phase, has to be large, as can be seen by the on-shell expression \eqref{eq:gravitationalsectorFLQP}.
\subsection{Comparison to Hartle-Hawking} \label{subsec:comparisonHHbraket}
Until now we have discussed only the connected geometry contribution to the Wigner distribution. There is also a disconnected geometry that contributes, which is enhanced by a relative factor $e^{2 \phi_0}$. It is basically the Hartle-Hawking solution with modified boundary conditions.  We thus need to add up the contributions coming from two types of geometries and study the full distribution.The situation for the disconnected geometry is, in a sense, the opposite of that for the bra-ket wormholes. While the on-shell solution to the density matrix can be readily found, classical saddles for Wigner boundary conditions are rather subtle. It is easiest to first compute the density matrix and then do the transform.  We discuss this calculation in Appendix \ref{app:wignertransformofHH}, where we also briefly mention classical solutions. In the large-$L$ limit for the case where we use the inflaton as time we get, using  \eqref{eq:HHLQPWigner},
\begin{align}
\label{distrHH}
    W_{\text{HH}}(L,\Phi,U,V|f_b) = e^{2\phi_0} \frac{\Phi^2}{L^4}\delta(U-U_{\text{cl}}(L))\delta(V-V_{\text{cl}}(L))\,,
\end{align}
where the classical solutions, now parametrized by $L$, are given by:
\begin{align}
    U_{\text{cl}}(L)=2(1+ \frac{1}{2 L^2}) + O(\lambda), \qquad
    V_{\text{cl}}(L)=2(1-\frac{1}{2 L^2}) + O(\lambda). \qquad
\end{align}
Unlike in the bra-ket case, the $\delta$-functions do not get regularized by a finite $\lambda$, instead they get
smeared because  $L$ and $\Phi$ are finite. 

To compare the contributions of connected and disconnected geometries we can, as in section \ref{observer}, consider measurements of some observable localized in a region much smaller than $L$. We again assume that such observables cannot depend on $L$ or $\Phi$ directly, however, they depend on $U$ and $V$. We see that both terms in the distribution (\eqref{eq:wignersmallpq} and \eqref{distrHH}) project on classical solutions. These solutions are different at subleading order at late times, so an observer with enough precision would be able to distinguish the two. The relative weight of the two contributions is given by
\begin{align}
\label{WRatio}
\frac{W_{\text{bra-ket}}}{W_{\text{HH}}}\approx \frac{L^4}{\Phi^2} e^{-2\phi_0}\,.
\end{align}
We see that for large enough universes the bra-ket contribution will dominate. This effect is somewhat similar to the domination of the double-cone geometry responsible for a ramp in the spectral form-factor \cite{Saad:2018bqo}, see also \cite{Chen:2023hra} Note, however, also an important difference. Our connected contribution is enhanced with respect to disconnected by powers of geodesic length of the boundary, not by the renormalized length, or by factors of $L/\Phi$ only. This may appear surprising because the calculations appears very similar, given that spectral form factor also needs to be integrated over temperatures in order to correspond to an on-shell solution. Nevertheless, in the AdS discussion only renormalized quantities appear, as is evident from the boundary dual. The difference comes from the interpretation of the observables in dS and in AdS. Namely, we see that $W_{\text{HH}}$ is a narrowly-peaked function, almost a delta-function on the phase space, while $W_{\text{bra-ket}}$ has an almost flat direction $V$, which corresponds to the age of the universe. When evaluating the ratio \eqref{WRatio} we assumed that the range of integration in $V$ is of order one, thus we do not put any priors on the length of inflation, $\Delta N_e\sim N_e$. This is not completely consistent since it would include contributions of young universes. An extreme opposite limit is to assume that integration range of $V$ is of order $e^{-2N_e}$, or $\Delta N_e\sim 1$. In this case we are integrating the (smeared) delta function in the HH piece on the scales of order of its width, see appendix \ref{app:wignertransformofHH}. It still integrates approximately to 1, however, the integral of the smooth bra-ket contribution will be suppressed by $e^{-2N_e}$. This will roughly change the factor of $L^2$ in \eqref{WRatio} to the renormalized boundary length, more similar to the AdS result. We think that in cosmology a reasonable prescription would be to consider an intermediate situation $N_e\gg\Delta N_e\gg 1$, which corresponds to keeping the first term in \eqref{localVint}. Then the ratio in \eqref{WRatio} would change to 
\begin{align}
\frac{W_{\text{bra-ket}}}{W_{\text{HH}}}\approx \frac{L^2}{\Phi^2} e^{-2\phi_0} L ^2 e^{- 2 (N_e-\Delta N_e)}\,,
\end{align}
so that we get a factor of renormalized boundary length squared, enhanced by a large factor $e^{\Delta N_e}$.


If we add conformal matter, the bra-ket wormhole contribution will be additionally enhanced by the matter partition function. This factor can also dominate over the topological factor for large enough universes, as in \cite{Chen:2020tes}. In Appendix \ref{app:entanglemententropy} we discuss the entropy paradox formulated in  \cite{Chen:2020tes} and check that for very similar reasons the paradox is only present on the disconnected saddle, and that the bra-ket wormhole always starts to dominate before the paradox can occur, at least for the range of parameters and the region of classical phase space in which we have good control of the calculation.

\section{Temperature as Time}\label{sec:temperatureastime}
In previous sections we have considered different types of Wigner distributions and analysed their probabilistic interpretation with respect to observations motivated by the inflationary phase in cosmology. In particular, we used one of the scalar fields as a time variable. In this section we consider an alternative choice of time variable, inspired by a hot Big Bang period of the cosmological evolution. During this period there is no classically evolving elementary field that could be chosen as a clock field, instead, it is customary to choose the temperature of the thermalized matter component as time. Indeed, once primordial plasma cools down to a certain temperature, certain events happen: a given particle decouples, galaxies start to form, etc. The advantage of the time variable we use in this section is that for its large values the universe is automatically old, so this way we alleviate the age of the universe problem in our model. 

In the two-dimensional toy model of cosmology that we are considering, we can think of the CFT fields as a model of radiation filling in the expanding universe. The Euclidean evolution prepares a thermal state for the CFT, then the fields dilute and cool down during subsequent Lorentzian evolution resulting in a decrease of the temperature and therefore of the energy density of the CFT. In particular, the latter is given by the time component of the traceless stress tensor of the CFT\footnote{The trace term that we remove is the same that we reabsorbed in the definition of $\phi_0$ throughout the main text. Since we have fields that evolve with time and break dS isometries it is easy to write a covariant expression that will reduce to \eqref{eq:energydensitytraceless} on-shell. }
\begin{align}\label{eq:energydensitytraceless}
	\varepsilon=T_{tt}-\frac{g_{tt}}{2}  T^\mu_{\mu}\,.
\end{align}

Along the lines of the previous section, we consider a local observer making measurements at a certain moment in the evolution of the Universe, with the difference that we imagine such an observer being endowed with a measuring device, a clock $\mathcal{T}$, that is coupled to the CFT in such a way that it can measure its temperature, and therefore its energy density $\varepsilon$.

We consider the system introduced at the end of section \ref{sec:additionofinflaton} with matter consisting of both CFT and inflaton,\footnote{We still keep equal Dirichlet boundary conditions on the inflaton for simplicity, even though we are not using it as time.} $\mathcal{T}_K=\mathcal{T}_B=\mathcal{T}_b$. In the solutions we found, the energy density $\varepsilon$ is different at bra and ket sides, however, we will work in the classical regime where the opening of the contour is small, see section \ref{subsec:interpretation_classicality}. 
We thus can use a simple model for the clock which amounts to adding the following (boundary) term to our action:
\begin{align}
    S_\mathcal{T}= -\frac{\gamma}{2 \pi} \int d \varphi\left(\mathcal{T}_b+ E\right)^2\,,
\end{align}
where $E=\varepsilon_K+\varepsilon_B$ is the sum of the energy density \eqref{eq:energydensitytraceless} at bra and ket and $\gamma$ is a large parameter.  Then the path integral will be strongly dominated by configurations such that $E=-\mathcal{T}_b$ at the boundary.

Let us determine the distribution with this new time variable. The total action is
\begin{align}
	iS=iS_{W(\Phi,L,P,Q)}+iS_f+ \log(Z_{\text{CFT}})+S_\mathcal{T}\,.
\end{align}
Following the approximation we used above, the on-shell effective value of the energy density \eqref{eq:energydensitytraceless} is given by the Casimir energy and conformal anomaly (in FLRW Milne coordinates \eqref{eq:milnemetric}):
\begin{align} \label{eq:energydensityonshell}
    \varepsilon(t)=\frac{1}{a(t)^2}\frac{c \pi}{6}\left(\frac{1}{\beta^2}-\frac{1}{4 \pi^2}\right)\,,
\end{align}
where we isolated the scale factor $a(t)^2= \sinh^2 t$. Its sum at ket and bra becomes
\begin{align}
    E=\varepsilon_K+\varepsilon_B=\frac{c e^{-2 t_c} \cos \left(2 t_E\right)}{\pi }+ \mathcal{O}\left(e^{-4t_c}\right)\,,
\end{align}
where we used the expansion \eqref{eq:temperature} for $\beta$. As we explained after reintroducing units in \eqref{eq:wignerFLPQwithCFTandunits}, the CFT contribution can be subleading because it is a one-loop effect, such that we will neglect its contribution to the action here. After substituting $t_c$ and $b$ obtained in \eqref{eq:gravitationalsectorFLQP}, the action as a function of $t_E$ is 
\begin{align}\label{eq:Tastimeoffshell}
    i S(t_E)&=  2\left(\Phi L-P Q\right)\tan (t_E)+ \pi  \lambda  \phi _0 (L+Q) \left(2 f_b+\lambda \right)\tan (t_E)-\nonumber\\
    &-4 \lambda ^2  \phi _0 \sqrt{Q^2-L^2}  (\pi -t_E) t_E \frac{1}{\cos t_E}+ \\
    &-\gamma  \left(\mathcal{T}_b+\frac{c (Q-L)}{\pi  (Q+L)}\cos(2t_E)\right)^2\nonumber\,.
\end{align}
Extremizing this gives (up to the normalization factor) 
\begin{align}
	&W(L,\Phi,Q,P; f_b| \mathcal{T}_b)=\exp\left(-\gamma \left(\mathcal{T}_b+\frac{c (Q-L)}{\pi  (Q+L)}\right)^2\right)\\
	&\times\exp\left(-\frac{\left(2(\Phi L-P Q)+\pi  \lambda  \phi _0 (L+Q) \left(2 f_b+\lambda \right)\right){}^2}{16 \lambda ^2 \phi _0 \sqrt{Q^2-L^2}}\right)\nonumber\,,
\end{align}
where we assumed that the solution for $t_E$ is not affected by the clock term in the action.
Notice that the first factor in this distribution for large values of $\gamma$ becomes a delta function that enforces the following relation between $Q,L,\mathcal{T}_b$:
\begin{align}\label{eq:Tastimeconstraint}
	\mathcal{T}_b=-\frac{c (Q-L)}{\pi  (Q+L)}=-\frac{c \sqrt{V-1}}{\pi  \sqrt{V+1}}\,,
\end{align}
upon which the distribution becomes \eqref{eq:LFQPdistribution} or rewriting with $U,V$ variables \eqref{eq:wignersmallpq}. For large universes (large $\kappa\equiv \frac{\Phi \sqrt{L}}{\sqrt{16 \lambda^2 \phi_0}}$) the latter becomes a delta function, so that the full Wigner distribution with the temperature as time variable is
\begin{equation} \label{eq:Tempastimewignerdelta}
	W(\Phi,L,U,V; f_b\,|\mathcal{T}_b)\propto\frac{1}{V}\delta(U-U_{\text{cl}}(V)) \, \delta\left(\mathcal{T}_b+\frac{c \sqrt{V-1}}{\pi  \sqrt{V+1}}\right)\,.
\end{equation}
When integrating observables against this Wigner distribution as in \eqref{eq:expvalWignerUV}, the integral over $V$ now simply replaces it with a corresponding function of $\mathcal{T}_b$. In particular, at late times, this means that $\mathcal{T}_b$ approaches zero from below, and $V$ is forced to be close to 1, which corresponds to an old universe.

\section{Conclusions}\label{ref:summaryandoutlook}
In this work we found new saddles of the gravitational path integral in de Sitter JT gravity with various matter sources. The boundary conditions that we imposed correspond to the Wigner distribution -- a certain Fourier transform of the density matrix which is a function on a classical phase space. The resulting geometries have two Lorentzian regions that are naturally associated with bra and ket of the wave function of the universe, however, they are also connected by a region with complex metric. Thus these geometries correspond to the bra-ket wormholes of \cite{Chen:2020tes}. For a suitable choice of parameters, importantly for large enough universes, our connected contribution dominates over the disconnected one, the latter being a standard Hartle-Hawking saddle. The connected solution exists for pure JT gravity, thus the wormhole is stabilized not by matter, but rather by boundary conditions on the metric, as in the double-cone geometry \cite{Saad:2018bqo}. It seems like our solution has much in common with that geometry and it would be interesting to explore the connection further. Even though the solution exists without matter, it does not lead to a very interesting distribution, which in the simplest case is just a monotonic function.

For this reason we introduce matter and treat it as a perturbation. The solution still exists, but it gets slightly deformed and leads to a more interesting distribution. In particular, we add an inflaton scalar field with a linear potential and use it to fix the time diffeomorphisms. This produces a distribution that is narrowly peaked around classical solutions of the theory, see expression \eqref{eq:LFQPdistributionupgraded}. This is in accord with the expectations for the Wigner distribution of a generic semiclassical state. Ideally, the Wigner distribution should also give a probability measure on the integration constants of the equations of motion. This would fully solve the problem of determining the initial conditions for the universe at the semiclassical level.  Unfortunately, our distribution is not normalizable with respect to these parameters. We emphasize that these are not divergences in the path integral that computes the distribution. These are divergences that appear when studying the probabilistic properties of the distribution. One type of divergence is related to the limit of large universes. It could be that some non-perturbative effects become important in that limit. This is what happens in the spectral form factor example, for which the ramp transitions to a plateau due to such effects \cite{Cotler:2016fpe,Okuyama:2020ncd,Saad:2022kfe,Altland:2022xqx,Blommaert:2022lbh}. At the moment we do not have any proposal for how such effects can be calculated in a cosmological context. Another divergence is related to a parameter of classical solutions which we call the ``age of the universe''. It appears because for most of our choices of the time variable, large values of that variable do not necessarily imply a long Lorentzian evolution of the universe. In fact, unlike in the Hartle-Hawking case, an arbitrarily large universe can be created at an instant by a bra-ket wormhole. We do not have control over our geometry in that limit which also leads to a divergence. This can be thought of as a version of the ``short inflation'' problem of the higher-dimensional HH wave function, which was one of our motivations. In our simple model we found one way to address this problem -- we related the time variable to the temperature of the matter fields such that when this temperature is small, the universe is automatically old. It is possible that there are other ways to address this important problem using bra-ket wormholes. In particular, it would be interesting to study bra-ket solutions in modified JT gravity recently discussed in \cite{Blommaert:2024ydx,Collier:2024kmo} as well as in other models of two-dimensional cosmology \cite{Anninos:2024iwf} and see if they have better normalizability properties. Another source  of corrections to dS JT gravity appear from
reduction from higher dimensions. For example, in \cite{Maldacena:2019cbz} it was observed that such corrections improve normalizability of the HH state. It would be nice to see how similar corrections affect normalizability of the Wigner distribution. 

The last point brings us to the most exciting future direction, which is to construct a bra-ket wormhole solution in a four-dimensional inflationary model and see if it produces a distribution consistent with the observations in the real universe.  Because the JT gravity model we considered is a reduction of 4D gravity, one can be optimistic that our solution can be lifted to 4D. In this case, one gets an $S_2\times S_1$ or $S_2\times \mathbf{R}$ spatial topology. We can also consider other topologies in addition to the standard $S^3$, for example, various compact and non-compact hyperbolic manifolds. Some recent work on cosmological wormholes can be found in \cite{Aguilar-Gutierrez:2023ril, Aguilar-Gutierrez:2023hls}. Double-cone wormholes, to which our solution appears analogous, are expected to exist in arbitrary dimensions in generic theories of gravity and to be enhanced by the universal ramp-like effect \cite{Saad:2018bqo,Chen:2023hra}. As we mentioned above, bra-ket wormholes can right away produce a very large or non-compact universe, with a homogeneous state of matter on it. It thus can provide an alternative solution to the so-called horizon problem, even without a period of long inflation. One can speculate what kind of phenomenological predictions a model of this sort can have. Generically, we expect some form of anisotropy at large scales, related to a spatial topology to appear, as well as corrections to the power spectrum, also at long distances, in the spirit of the formula \eqref{eq:scalartwopointphysical}. Additional corrections to correlation functions can result from the bra-ket wormholes of the type considered in reference \cite{Mirbabayi:2023vgl}. As a step towards a realistic construction, one can also consider three-dimensional gravity, where both the choice of spatial topology and quantum corrections are more tightly under control. 

\subsection*{Acknowledgements}
We thank Yiming Chen, Jan de Boer, Luca Iliesiu, Oliver Janssen, Juan Maldacena, Mehrdad Mirbabayi, Veronica Sacchi, Julian Sonner, Mattia Varrone, and Zhenbin Yang for discussions and comments. 
JKK is
supported by the National Centre of Competence in Research SwissMAP. VG acknowledges support form the SNF starting grant ``The Fundamental Description of the Expanding Universe''.
\appendix
\section{Hartle-Hawking Density Matrix and its Wigner Transform}\label{app:wignertransformofHH}
In this section we first review the Hartle-Hawking wave function with Schwarzian corrections and then consider the Wigner transforms of the corresponding density matrix.
The Hartle-Hawking geometry is discussed in global coordinates, equation \eqref{eq:globalmetric}. The prescription demands the joining of the Lorentzian geometry \eqref{eq:globalmetric} at the point $\hat{t}=0$ to a continuation of that same geometry to the Euclidean half-sphere:
\begin{equation}
ds^2= d \theta^2 + \cos(\theta)^2 d \hat{\varphi}^2\,,\quad 0 \leq \theta \leq \pi/2\,,
\end{equation}
via $\hat{t}= i \theta$. The contour on the complex $\hat{t}$ plane is shown in figure \ref{fig:HHcontour1}. Note that we refer to this contour as Hartle-Hawking contour, in spite of the fact that in the original proposal the wave function contained two contributions with the opposite singes of the phase. As before the on-shell action of the dynamical bulk term is zero. The topological term along the contour shown in figure \ref{fig:HHcontour1} can be split into two:
\begin{equation}
  i S_{\text{top.}} = i  \phi_0\left(\int_{\hat{t}=0}^{\hat{t}=\hat{t}_c} d\hat{t} \cosh(\hat{t}) - \sinh(\hat{t})\Big|^{\hat{t}=\hat{t}_c}_{\hat{t}=0}\right) + i \phi_0 \left( \int^{0}_{\pi/2}i d\theta \cos (\theta)\right) = \phi_0\,,
\end{equation}
where the first term corresponds to the Lorentzian section and the second term to the Euclidean half-sphere. 
In addition the boundary term supplies 
\begin{equation}
    i S_{\text{bdy}}=\left( -  i\phi_b \ell_K + \frac{i\phi_b}{2 \ell_K}\right)\,.
\end{equation}
in late time expansion. This can be extended to include Schwarzian fluctuations resulting in the expression \cite{Maldacena:2019cbz}:
\begin{equation}\label{eq:wavefunctionHHexplicitly}
    \Psi_{HH}(\ell_K|\phi_b)=\left(\frac{\phi_b}{\ell_K}\right)^{3/2}\exp\left( \phi_0   - i \phi_b \ell_K + i \frac{\phi_b}{2 \ell_K} \right)\,.
\end{equation}
Up to a convention dependent number.
\begin{figure}[]
\centering
\subfigure[]{ 
\label{fig:HHcontour1}
\includegraphics[width=0.3\textwidth]{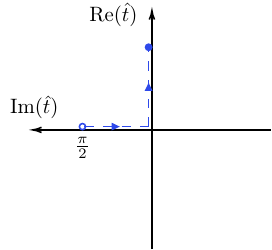}}
    \hspace{70pt}
    \subfigure[]{
    \label{fig:HHcontour2}
\includegraphics[width=0.3\textwidth]{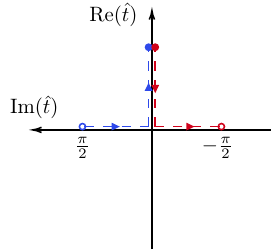}
    }
    \caption{(a) The integration contour of the Hartle-Hawking wave function in terms of the global coordinates defined in \eqref{eq:globalmetric}. (b) The contour for the Hartle-Hawking density matrix. Bra and ket are shown in blue and red respectively. The bra and ket future boundaries are denoted by blue and red dots, the poles of the half-spheres are instead denoted by circles. Note that there is no connection between bra and ket in the past. On the bra side time goes in the opposite direction because of complex conjugation. For the Wigner distribution of this quantity we integrate over the full domain of possible differences between bra and ket contour endpoints.}
\end{figure}
We give some more details on Schwarzian corrections in Appendix \ref{app:Schwarziancorrections}.
The first term in the exponential comes from the topological term, whereas the other two come from the boundary term of \eqref{eq:actionfordirichlet}. The above HH-wave function \eqref{eq:wavefunctionHHexplicitly} in turn defines a factorisable contribution to the density matrix of the universe \eqref{eq:formalexpressiondensitymatrix}:
\begin{equation}
    \rho_{\text{HH}}(\ell_K,\ell_B|\phi_b) = \Psi^{*}(\ell_B,\phi_b)\Psi(\ell_K,\phi_b)\,,
\end{equation}
which explicitly amounts to
\begin{equation}\label{eq:densitymatrixHH}
    \rho_{\text{HH}}(\ell_K, \ell_B|\phi_b)= \frac{\phi_b^3}{(4 L ^2-\Delta \ell^2)^{3/2}}\text{exp}\left( 2 \phi_0  -  i  \phi_b \Delta \ell + i  \phi_b \frac{2 \Delta \ell}{(\Delta \ell)^2 -  4 L^2}\right)\,.
\end{equation}
The diagonal elements merely consist of the topological term. Let us first note that this is a Hermitian density matrix: $\rho(\ell_K,\ell_B)=\rho(\ell_B,\ell_K)^{\ast}$.
We would like to perform the Wigner transform of this expression, i.e perform the integration
\begin{equation}\label{eq:formalexpressionWignertransformHH}
    W_{\text{HH}}(L,P|\phi_b) = \int_{-2 L_\Lambda }^{2 L_\Lambda }d \Delta \ell \,\rho_{\text{HH}}(\ell_K, \ell_B|\phi_b) e^{i P \Delta \ell}\,.
\end{equation}
Here we have restricted the integration range as we are considering $\ell_K, \ell_B\gg 1$, because outside of that range we need to include finite cutoff corrections.\footnote{See reference \cite{Iliesiu:2020zld,Stanford:2020qhm} for work on the JT gravity at finite cutoff.} Clearly the result will still exhibit the topological weighting present in \eqref{eq:wavefunctionHHexplicitly} and \eqref{eq:densitymatrixHH}. Next we want to understand the functional behaviour of \eqref{eq:formalexpressionWignertransformHH} in terms of $P$ and $L$. 
 Let us assume that the integral is well-approximated by $L_\Lambda << L$. Hence we expand the last term in the exponential of the integrand for small $\Delta \ell$.
At leading order this leads to a linear term in $\Delta \ell$ with a correction in form of a cubic term, which we will drop for the moment. Then inside the integral $\Delta \ell$ acts as a Lagrange multiplier. Setting a cutoff $L_{\Lambda}$ on the integration boundaries leads to a regularised $\delta$-function of width $1/L_{\Lambda}$. Taking it large therefore leads to the result
\begin{equation}\label{eq:HHLPWigner}
    W_{\text{HH}}(L,P|\phi_b) \sim e^{2\phi_0}\frac{\phi_b^3}{L^3} \delta\left(\phi_b(1+\frac{1}{2 L^2})-P\right)\,.
\end{equation}
As can be checked by the explicit equations of motion, the $\delta$-function peak is at the on-shell value of the variables $L$ and $P$.
Now if we include the cubic term, we can find two saddlepoints which lie at 
\begin{equation}\label{eq:saddlepointHH}
    \Delta \ell = \pm \frac{2 L \sqrt{2 L^2(P-\phi_b)-4\phi_b}}{ \sqrt{3}\sqrt{\phi_b}}\,,
\end{equation}
and are therefore inside the allowed region. These saddles  can be also found semi-classically by directly solving the bulk equations of motion with Wigner boundary conditions. The saddlepoints in \eqref{eq:saddlepointHH} also lead to a regularized expression for the $\delta$-function of the on-shell values of phase-space variables, but with the width given by $\frac{1}{L}\left(\frac{\phi_b}{L}\right)^{1/3}$, which is also small in the large-$L$ limit. We can trust the saddle points in the limit $\frac{\phi_b}{L} >> 1 $. This is also the limit in which we can trust the bulk solutions, as long as the phase space variables are within the peak of the smeared $\delta$-function. We can drop the cubic in $ \Delta \ell$ term in the opposite limit $\frac{\phi_b}{L} \ll 1$ and in this limit we do not find any bulk solution. In any case we get a $\delta$-function as in \eqref{eq:HHLPWigner} up to an order-one constant and corrections that vanish when $L$ is large.

We can also determine the Hartle-Hawking Wigner distribution of the type \eqref{eq:formalexpressionsecondWignerfunction} which we denote with $W_{\text{HH}}(L,Q,\Phi,P|f_b)$. For this we have to start with a slightly more general density matrix: 
\begin{align}
    &\rho_{HH}(\phi_K,\phi_B,\ell_K,\ell_B|f_b) \nonumber\\
    &=\text{exp}(2\phi_0)\left(\frac{4 \Phi^2-\Delta\phi^2}{4 L^2-\Delta \ell^2}\right)^{3/2}\text{exp}\left(-i\left(\Delta \ell \Phi + \Delta\phi L \right) + \frac{2 i \left(\Delta \ell\ (\Phi )- \Delta \phi L \right)}{\Delta \ell^2 - 4 L^2}\right)\,.
\end{align}
Even though we have dropped $\lambda$ corrections in this expression one should still consider the inflaton to act as the time variable here. We comment below on how the result is affected by explicitly including the inflaton contributions. 
The integral transform we have to perform is 
\begin{equation}\label{eq:HHdilatonastimeWigner}
    W_{\text{HH}}(L,Q,\Phi,P|f_b)=\int_{-2 L}^{2 L} d \Delta \ell\int_{-2\Phi }^{2\Phi }d \Delta\phi \rho_{\text{HH}}(\phi_K,\phi_B,\ell_K,\ell_B) e^{i P \Delta \ell} e^{i Q \Delta \phi}\,.
\end{equation}
We consider similar logic as for the previous transform. For both integrations we do not have control over the full range as our results for the density matrices only hold in the regime of $\phi_b$ and $\ell$ large. The $\Delta \phi$ integral then supplies an extra $\delta$-function, this time regularized by the $\Phi$ integration range. Overall this results in the expression
\begin{equation}\label{eq:HHLQPWigner}
    W_{\text{HH}}(L,Q,P,\Phi|f_b) \sim\text{exp}(2\phi_0)\left(\frac{\Phi^3}{ L^3}\right)\delta\left(\Phi(1+ \frac{1}{2 L^2})-P\right)\delta\left(L(1-\frac{1}{2L^2})-Q\right)\,.
\end{equation}
 In the approximation we are working in, the inflaton only backreacts on the dilaton. Hence, both dilaton variables are changed via backreaction. In the above expressions this merely shifts the peak of the delta functions in a $\lambda$-dependent manner. In addition there is an Euclidean contribution which shifts $2\phi_0$ in the exponential by a small amount.
\section{Gravitational Actions and Boundary Conditions}\label{secapp:gravitationalaction}
In this section we determine the appropriate actions for the three types of Wigner functions we have discussed in the main text, namely the functional integrals \eqref{eq:formalexpressionwignertransform}, \eqref{eq:formalexpressionsecondWignerfunction} and \eqref{eq:formalexpressiontemperatureWignerfunction}. Discussion of boundary conditions in the context of JT gravity can be found in the references \cite{Cvetic:2016eiv,Blommaert:2019wfy,Ferrari:2020yon,Goel:2020yxl}. For convenience let us do this explicitly in FLRW coordinates, which are of the form
\begin{align}
    ds^2=-dt^2+a(t)^2d\varphi^2\,\quad \varphi \sim \varphi + 2 \pi b\,,
\end{align}
where choosing $a(t)=\sinh(t)$ gives Milne coordinates, equation \eqref{eq:milnemetric}.
Let us start by considering the dynamical bulk term of \eqref{eq:Jtbulklagrangian}. In FLRW and minisuperspace approximation this amounts to:
\begin{equation}
    S= \frac{1}{2\pi} \int d\varphi dt \left(\Ddot{a}-a\right)\phi \,.
\end{equation}
The dynamical fields here are the scale factor $a$ and the dilaton. By taking the variation with respect to $\phi$ one trivially gets the equations of motions for the metric, without any need of adding counterterms to ensure the validity of the variational principle. Taking the variation with respect to $a$ instead gives after performing partial integration
\begin{align}\label{eq:boundaryvariation}
    \delta_a S &= \frac{1}{2 \pi}\int dt d \varphi (\Ddot{\phi}-\phi) \delta(a) +\nonumber\\
    &+\frac{1}{2\pi}\int d\varphi \left( \left(\delta (\Delta \dot{a})\, \Phi + \delta (\dot{\mathcal{A}})\,\Delta \phi\right) -\left(\delta (\Delta a) \,\dot \Phi + \delta (\mathcal{A})\Delta \dot{\phi}\right)\right)\,.
\end{align}
The boundary terms are written in terms of sums and difference of boundary quantities with uppercase letters indicating sums and $\Delta$ indicating differences: $X\equiv \frac{x_K+x_B}{2}$ and $\Delta x\equiv x_K-x_B$. For a single boundary the different possible boundary conditions are determined by the dilaton and the boundary metric and their canonical momenta, amounting to the normal derivative of the boundary metric and the normal derivative of the dilaton. Here we see that the presence of two boundaries furnishes linear combinations. These quantities are obviously related to the boundary length, its normal derivative (we denote with $Q$ its sum at bra and ket as in \eqref{eq:Qdefinition}), the dilaton, and its normal derivative:
\begin{align}
    L= \frac{1}{2 \pi}\int d \varphi \mathcal{A}\,, \quad& Q= \frac{1}{2 \pi}\int d \varphi \Dot{\mathcal{A}}\, .
\end{align}
There is a subtlety in the definition of the above bra and ket quantities for the $\pi$-contour and deformations thereof, that lies in the branch of the induced volume form to be chosen, i.e. the sign of $\sqrt{h}_{K,B}=\pm a(t_{K,B})$.
Our choice is to pick the sign for which the real parts of the various quantities are positive at both bra and ket, consistently with what we have done for the boundary lengths in
\eqref{eq:boundarylengthdefinition} and gives:
\begin{align}
     \mathcal{A}= \frac{a_K-a_B}{2} \,, \quad &\dot{\mathcal{A}}= \frac{\dot {a}_K+\dot {a}_B}{2}\,\nonumber\\
      \Delta a= a_K+a_B \,, \quad &\Delta(\dot{a})= \dot {a}_K-\dot {a}_B\,.
\end{align}
where we also take into account the change of signs of the normal derivatives at the boundaries. For the quantities involving the dilaton instead the ambiguity is not present and we have
\begin{align}
     \Phi= \frac{\phi_K+\phi_B}{2}\,, \quad &P= \Dot{\Phi}=\frac{\dot{\phi}_K-\dot{\phi}_B}{2}\,, \nonumber\\
    \Delta\phi= \phi_K-\phi_B\,, \quad &\Delta \dot{\phi}=\dot{\phi}_K+\dot{\phi}_B\,, 
\end{align}
From the expression \eqref{eq:boundaryvariation} we can read off the dual momenta involved in the Wigner transforms $\Delta \ell \leftrightarrow P$ and  $\Delta \phi \leftrightarrow Q$, and identify which counterterms one needs to add to the action to ensure the validity of the variational principle for each choice of boundary conditions. We recognise the first two terms as the standard Gibbons-Hawking term.
In sections \ref{sec:bra-ketwormholes} and \ref{sec:braketwhandconformalmatter} we consider the dilaton $\phi$ to function as our clock field, such that we set Dirichlet conditions on it: $\Phi= \phi_b\,,\, \Delta \phi=0$. 
We can also see that the third term of \eqref{eq:boundaryvariation} comes into play when we do not fix the boundary length difference. 
Therefore the correct action for the Wigner distribution $W(L,P|\phi)$ is given by
\begin{equation}\label{eq:FLRWactionW(L,P,phib)}
S_{W(L,P|\phi_b)}=\frac{1}{2\pi} \int d\varphi dt \left(\Ddot{a}-a\right)\phi+\frac{1}{2\pi}\int d\varphi \left( \Delta a P -  \Delta \dot{a} \phi_b \right)=S_{DD} +  \Delta \ell \, P\,.
\end{equation}
where we used $P=\dot{\Phi}$. This is the same as expression \eqref{eq:boundaryactionWigner}. In section \ref{sec:additionofinflaton} we consider the addition of an inflaton field $f$. As explained in the main text we therefore consider Dirichlet conditions for $f$ and free up the the difference in the dilaton values, i.e only fix $\Phi$ at the boundary. From \ref{eq:boundaryvariation} we can read off that this implies a boundary condition on the conjugate momentum $Q\sim \int d\varphi \Dot{\mathcal{A}}$. For the JT sector therefore the appropriate action for $W(L ,Q,\Phi,P|f_b)$ is 
\begin{equation}\label{eq:actionforinflatonastime}
    S_{W(L ,Q,\Phi,P)}=\frac{1}{2\pi} \int d\varphi dt \left(\Ddot{a}-a\right)\phi   +\frac{1}{2\pi}\int d\varphi \,  \left(\Delta a P - \Delta \dot{a}\Phi\right)=S_{DD} + \Delta \ell \, P +  \Delta \phi\, Q\,,
\end{equation}
which of course is supplemented by the inflaton action. In section \ref{sec:temperatureastime} we also encounter the case where we set Dirichlet conditions on the scale factor but Wigner conditions on the dilaton, i.e. fixing $(\Phi,Q)$, so that the arguments of the Wigner distribution are $W(\Phi,Q|\ell_b)$ looking at \eqref{eq:boundaryvariation} we see that the appropriate action is
\begin{align}\label{eq:action_dirichlet_a}
    S_{W(\Phi,Q|\ell_b)}=\frac{1}{2\pi} \int d\varphi dt \left(\Ddot{a}-a\right)\phi-\frac{1}{2\pi}\int d\varphi \,  \Delta \dot{a}\,\Phi=S_{DD} + \Delta \phi\, Q\,.
\end{align}
\section{Schwarzian Corrections}\label{app:Schwarziancorrections}
In this appendix we display some details regarding the integral over Schwarzian fluctuations. This is relevant to both the Hartle-Hawking wave function \eqref{eq:wavefunctionHHexplicitly} and the prefactor of the off-shell density matrices used in section \ref{sec:oneloopdeterminant} and the Appendix \ref{app:inverseWigner}. Schwarzian Corrections correspond to introducing a ``wiggly'' boundary under the constraint of a fixed geodesic length \cite{Engelsoy:2016xyb,Maldacena:2016upp}. Let us start with Hartle-Hawking wave function. In the global coordinate system \eqref{eq:globalmetric}, we set the following boundary condition:
\begin{equation}
\cosh(\hat{t}_K(u))\hat\varphi_K'=\ell_K\,.
\end{equation}
Here $u$ is the intrinsic boundary coordinate and the $'$ here denotes a derivative with respect to that coordinate. Solving for $\hat{t}_K(u)$ as an expansion in large $\ell_K$ and performing a partial integration leaves us with an action of the type 
\begin{equation}
i S = \phi_0 - i \phi_b \ell_K  + \frac{i\phi_b}{4 \pi \ell_K}\int_{0}^{2 \pi}du \left(\hat\varphi_K'^2- \left(\frac{\hat\varphi_K''}{\hat\varphi_K'}\right)^2\right)\,.
\end{equation}
The field $\hat{\varphi}_K$ is restricted to be monotone increasing and to wind once around the circle parametrized by $u$: $\hat{\varphi}_K(u+ 2 \pi) = \hat{\varphi}_K(u) + 2\pi$.
As the Schwarzian action is $SL(2,\mathbb{R})$ invariant we have to mod out by this symmetry when considering the path integral. We integrate over the symplectic manifold $\text{diff}(S^1)/SL(2,\mathbb{R})$, which leaves us with the following expression for the path integral \cite{Saad:2019lba}
\begin{equation}
\Psi_{\text{HH}}(\ell_K|\phi_b)=e^{\phi_0}e^{- i \phi_b \ell_K  }\int \frac{D \hat\varphi_K}{SL(2,\mathbb{R})}e^{\frac{i\phi_b}{4 \pi \ell_K}\int_{0}^{2 \pi}du \left(\hat\varphi_K'^2- \left(\frac{\hat\varphi_K''}{\hat\varphi_K'}\right)^2\right)}\,.
\end{equation}
The integration measure follows from the symplectic form of the manifold $\text{diff}(S^1)/SL(2,\mathbb{R})$ \cite{Alekseev:1988ce,Stanford:2017thb,Witten:1987ty}. As we are interested in perturbations of our saddle, let us expand $\hat{\varphi}_K=u + \epsilon(u)$. The aforementioned measure then takes on the form \cite{Saad:2019lba,Kirillov}
\begin{equation}\label{eq:Schwarzmeasurefluctuations}
\Omega=\frac{\alpha}{2}\int_{0}^{2\pi} du\left(d \epsilon'(u) \wedge d \epsilon''(u) - 2 \,\text{Schw}(u)d \epsilon(u)\wedge d \epsilon'(u)\right)\,,
\end{equation} 
where $\alpha$ is a convention dependent number. Decomposing $\epsilon$ into Fourier modes and discarding the zero modes leaves us with the expression
\begin{equation}\label{eq:HHSchwarzianpathintegral}
\Psi_{\text{HH}}(\ell_K|\phi_b)=e^{\phi_0}e^{- i \phi_b \ell_K  + \frac{i \phi_b}{2 \ell_K}}\prod_{n \geq 2} 4\pi \alpha \left(n^3-n\right)\int d \epsilon_n^{R} d \epsilon_n^{I}e^{- i\frac{\phi_b}{\ell_K}(n^4-n^2)((\epsilon_n^{R})^2 + (\epsilon_{n}^{I})^2)}
\,.
\end{equation}
We see that to guarantee convergence we must consider a slight imaginary deformation of the contour. Regularizing via the $\zeta$-function gives expression \eqref{eq:wavefunctionHHexplicitly}, where we have fixed $\alpha$ to avoid numerical factors. 

Let us now come to the density matrices of section \ref{sec:oneloopdeterminant}. Compared to Hartle-Hawking, there are now two boundaries and there is an additional internal modulus $b$, which breaks the $SL(2,\mathbb{R})$ symmetry down to $U(1)$. For both bra and ket we set Schwarzian boundary conditions as in the main text, equation \eqref{eq:Schwarzianbcmilne}. This induces fluctuations on both boundaries. For the ket we end up with an integral over fluctuations of the form \cite{Saad:2019lba}
\begin{equation}
e^{- i \phi_b \ell_K  }\int \frac{D \varphi_K}{U(1)}e^{\frac{i \phi_b}{4\pi \ell_K}\int^{2\pi}_{0} du\left( - b^2 \varphi_K'^2 -  \left(\frac{\varphi''_K}{\varphi'_K}\right)^2\right)}\,.
\end{equation}
We again consider the measure \eqref{eq:Schwarzmeasurefluctuations} but in the mode expansion of $\epsilon$ we only have to remove a single zero mode, such that we get an extra multiplicative factor compared to \eqref{eq:HHSchwarzianpathintegral} and hence a different power for the final result. The bra  amounts to the above expression with $\ell_K$ replaced by $-\ell_B$ and the overall off-shell density matrix is a product of the two, i.e expression \eqref{eq:densitymatrix1}. We can repeat the same steps but with the dilaton taking on different values for each boundary resulting in expression \eqref{eq:densitymatrix2}.
\section{Some Details on Inverse Wigner Transforms}\label{app:inverseWigner}
In this appendix we give some additional details on various inverse Wigner transforms. Our goal is to show that the exponentials of the off-diagonal elements starting from expression \eqref{eq:LFQPdistribution} can be matched to the direct approach from the transform of an off-shell density matrix (in the $\lambda$ to zero limit). This then gives us further justification to match the prefactors for the diagonal elements as we do in section \ref{sec:oneloopdeterminant}. We will not consider the more involved calculation of the prefactors away from the diagonal limit. Let us first come to the object $W(L,P;\phi_K,\phi_B|f_b)$, which can be defined via an inverse transform as in the main text, equation \eqref{eq:LPphifbwignerinverse}, which we restate here for convenience
\begin{align}
    W(L,P;\phi_K,\phi_B|f_b)=\int_{ L}^{\infty}dQ \, W( L,Q,\Phi,P|f_b) e^{- i Q \Delta \phi}\,.
\end{align}
This object amounts to a Wigner function in phase space variables  $P$ and $ L$. However in the variables $\phi_K$, $\phi_B$ it behaves like a density matrix. Hence, it is Hermitian in those variables. Performing a saddlepoint integration, leads to the saddlepoint \eqref{eq:Qsaddlepointforinversetransform}, which we also restate here for convenience
\begin{equation}
Q^{\ast}= \frac{\Phi  L}{P}+\mathcal{O}(\lambda)\,.
\end{equation}
The conjugate variable $\Delta \phi$ appears at order $\mathcal{O}(\lambda^2)$ in the above expression. The value of the action on the saddlepoint then amounts to
\begin{align}
&W(L,P;\phi_K,\phi_B|f_b)\nonumber\\=&\mathcal{N}_{ L,P,\phi_K,\phi_B}\text{exp}\left(- 2 i \Delta \phi  L + i \frac{\Delta \phi  LP}{ \Phi} - \frac{\lambda^2\phi_0\Delta\phi^2  L\sqrt{(\Phi^2-P^2)}}{P^3} \right) \label{eq:Wignerqfloffdiagonalviainversetransform}\,.
\end{align}
Here we have dropped $\mathcal{O}(\lambda)$ imaginary terms but included the first real contribution. We can see that the off-diagonal element exhibits a non-zero action, which is Hermitian as expected. Here we do not give an explicit expression for the prefactor and hence just denote it as $\mathcal{N}_{ L,P,\phi_K,\phi_B}$. It is a product of Gaussian fluctuations and the prefactor $\mathcal{N}_{Q,\Phi,L,P}$, which we determined in section \ref{sec:oneloopdeterminant} on the saddlepoint. We should arrive at the same result by considering the following transform starting from a density matrix:
\begin{align}
W(L,P;\phi_K,\phi_B|f_b)&=\int_{-2  L}^{ 2 L}(d\Delta \ell) \int_0^{\infty} b db \,\rho^{(b)}(\ell_K,\ell_B,\phi_K,\phi_B|f_b)e^{ i\Delta \ell P  }\,.
\end{align}
Here we have introduced a new type of density matrix
\begin{equation}\label{eq:densitymatrix2}
    \rho^{(b)}(\ell_K,\ell_B,\phi_K,\phi_B|f_b)=\frac{1}{2}\sqrt{\frac{4\Phi^2-\Delta\phi^2}{ 4 L^2-\Delta \ell^2}}\text{exp}\left(-i\left(\Delta \ell \Phi+ \Delta \phi  L\right) + 2 i b^2 \left(\frac{ L \Delta \phi  -\Delta \ell \Phi }{\Delta \ell^2 - 4 L^2}\right)\right)\,.
\end{equation}
We denote the inflaton field here as the clock field.
Even though we will neglect $\lambda$-corrections here in the following, they could easily be included. If we now perform the integration, we see that the saddlepoint equations are solved by 
\begin{align}
\Delta \ell &= \Delta \phi \frac{ L}{\Phi}\,,\\
b&= \pm \frac{L \sqrt{(\Phi-P)}\sqrt{4\Phi^2-\Delta \phi^2}}{\sqrt{2}\Phi^{3/2}}\,.
\end{align}
Naturally, as we are interested in the saddle appropriate for $b>0$ and real
we consider $ \Phi >\Delta \phi$. and $\Phi>P$, which leaves us with one saddle, which gives the exponential
\begin{align}
W( L,P;\phi_K,\phi_B|f_b) &= \mathcal{N}_{P, L,\phi_K,\phi_B}\text{exp}(- 2 i \Delta\phi  L + i  \frac{\Delta \phi  L P}{\Phi} )\,.
\end{align}
Comparing to \eqref{eq:Wignerqfloffdiagonalviainversetransform} we see that action agrees in $\lambda$ to zero limit. Now let us come to $W(Q,\Phi;\ell_K,\ell_B|f_b)$, which we also defined via an inverse transform in the main text, namely equation \eqref{eq:QFLbwignerinverse}, which we recall
\begin{equation}
    W(Q,\Phi;\ell_K,\ell_B|f_b)= \int_{0}^{\Phi} dP  \,W( L,Q,\Phi,P|f_b)e^{-i P \Delta \ell} \,.
\end{equation}
The saddlepoint as in the main text, formula \eqref{eq:saddlepointvaluePforinversetrafo}, is of the form
\begin{equation}
    P^{\ast}=\frac{\Phi L}{Q} + \mathcal{O}(\lambda)\,.
\end{equation}
This results in 
\begin{align}\label{eq:WignerQfoffdiagonalL}
    &W(Q,\Phi;\ell_K,\ell_B|f_b)\nonumber\\=& \mathcal{N}_{Q,\Phi,\ell_K,\ell_B}\text{exp}\left(- 2 i \Delta \ell \Phi + i \frac{\Delta \ell \Phi Q}{  L} -  \frac{\lambda^2 \phi_0\Delta \ell^2\sqrt{(Q+ L)(Q- L)}}{Q^2}\right)\,,
\end{align}
which again is Hermitian with respect to ket and bra boundary lengths. We have dropped an $\mathcal{O}(\lambda)$ imaginary term but kept the first real contribution at $\mathcal{O}(\lambda^2)$. Starting from the general density matrix, equation \eqref{eq:densitymatrix2}, we can perform the following transform 
\begin{align}
W(Q,\Phi;\ell_K,\ell_B|f_b)&=\int_{-2 \Phi}^{2 \Phi}(d\Delta \phi) \int_0^{\infty} b db \,\rho^{(b)}(\ell_K,\ell_B,\phi_K,\phi_B;f_b)e^{ i \Delta \phi Q  }\,.
\end{align}
which results in 
\begin{equation}
 W(Q,\Phi;\ell_K,\ell_B|f_b)= \mathcal{N}_{Q,\Phi,\ell_K,\ell_B}  \text{exp}\left(- 2 i \Delta \ell \Phi + i \frac{\Delta \ell \Phi Q}{  L} \right)\,.
\end{equation}
We again see that in the $\lambda$ to zero limit that the action is the same as in \eqref{eq:WignerQfoffdiagonalL}. This provides further confidence in matching the prefactors in the diagonal limit as we do in section \ref{sec:oneloopdeterminant}.
\section{Dimensional Reduction from the Four-Dimensional Geometry: Restoring Units}\label{app:units}
The nearly $dS_{2}$ gravity system that we are considering can be obtained by a dimensional reduction of a four-dimensional geometry that is approximately $dS_{2} \times S^{2}$,  where $dS_{2}$ may have compact or non-compact spatial slices. Such a four-dimensional geometry can be obtained in the near-extremal limit of Schwarzschild-de Sitter black holes \cite{Maldacena:2019cbz}.\footnote{$dS_{2}$  JT gravity and deformations can be obtained also from a circle reduction of $dS_{3}$ \cite{Kames-King:2021etp,Narovlansky:2023lfz}.}
The extremal limit of a Schwarzschild-de Sitter black hole corresponds to the largest choice for the mass, such that the position of the cosmological horizon and the black-hole horizon coincide. 
This solution is referred to as the ``Nariai'' solution \cite{nariaiBH,Ginsparg:1982rs}. Close to the horizon the metric can be rewritten as $dS_{2}\times S^{2}$
\begin{align}\label{eq:4dmetric}
	ds^{2}=ds_{(2)}^2+R_{dS}^{2} d\Omega_{2}^{2}\,.
\end{align}
An approximately $dS_2 \times S^2$ metric can be written by replacing in  the  above $R_{dS}\to r=(R_{dS}+\delta)$ where $\delta$ depends on $dS_2$ coordinates and parameterizes the (small) deviations from $dS_2 \times S^2$.
The four-dimensional Einstein-Hilbert action is \footnote{We do not specify the boundary terms since we used different boundary conditions throughout the paper.}
\begin{align}\label{eq:4daction}
	S_{4d}=\frac{M_p^2}{2}\int d^4x \, \sqrt{-g_{(4)}} \left(R_{(4)}-\frac{2}{R_{dS}^{2}}\right)+\text{boundary terms}\,.
\end{align}
Dimensionally reducing this action, performing a Weyl rescaling $ds_{(2)}^2 \to ds_{(2)}^2 \frac{R_{dS}}{r}$ and expanding $r=R_{dS}(1+\delta)$ in small $\delta$ gives us
\begin{align}
	S_{2d}=&\frac{M_p^2\, R_{dS}^2}{2}\int d^2x \sqrt{-g_{(2)}}\, R_{(2)}+\frac{M_p^2\, R_{dS}^2}{2}\int d^2x \sqrt{-g_{(2)}}\, 2 \delta \left(R_{(2)} - \frac{2}{R_{dS}^{2}}\right)+\\
 &+\text{boundary terms}\nonumber\,,
\end{align}
which is the standard JT action \eqref{eq:Jtbulklagrangian} with
$\phi_0=2 \pi R_{dS}^2 M_p^2$ and $\phi= 2 \phi_0 \delta$. Even if $\phi \ll \phi_0$, the dilaton still scales with $\phi_0$ so that $\phi \sim M_p^2 R_{dS}^2$.
Let us look also at the inflaton action
\begin{align}
    S_{4,f}=\int d^4x \sqrt{-g_{(4)}}\,\left(\frac{(\partial f)^2}{2}-\lambda f\right).
\end{align}
Performing the same dimensional reduction (considering only the $s$-wave sector) and the Weyl rescaling we get
\begin{align}
    S_{2,f}=R_{dS}^2\int d^2x \, \sqrt{-g_{(2)}}\, \left(\frac{(\partial f)^2}{2}-\lambda f\right)\,,
\end{align}
where we also neglected effects of the deviations from exact $dS_2 \times S^2$ choosing $r=R_{dS}$, i.e. we neglected the coupling of the dilaton with the inflaton.
Notice that $\lambda \sim \text{length}^{-3}$.
Choosing the gauge \eqref{eq:4dmetric}, in minisuperspace approximation, and redefining the coordinates by stripping off the de Sitter length
\begin{align}
    ds^2=R_{dS}^2\left(-d\tilde{t}^2+a^2 d\tilde \varphi\right),
\end{align}
we see that the gravitational action becomes precisely the one we used throughout this paper
\begin{align}
    S_{2,\text{grav}}=\frac{\phi_0}{4\pi}\int d^2\tilde x\,  \tilde{R}+\frac{1}{4\pi}\int d^2\tilde x \,\phi \left(\frac{d^2a}{d\tilde t ^2} - a\right)+\text{boundary terms}.
\end{align}
So if we define the quantities $\phi_b,P$ etc. using the variables $\tilde t$, $\tilde \varphi$ we get the same answers as in the previous sections. 
The inflaton action becomes
\begin{align}
    S_{2,f}=R_{dS}^2\int d^2x \, a \, \left(\frac{1}{2}\left(\frac {df}{d \tilde t} \right)^2-\lambda R_{dS}^2 f\right).
\end{align}
We therefore get the action \eqref{eq:action_inflaton_minisuperspace} we used throughout this work but with the substitution $\lambda \to \lambda R_{dS}^2$ and $\phi_0\to R_{dS}^2$. It is now straightforward to reintroduce the units in the various results. The Wigner distribution with CFT matter, i.e section \ref{sec:braketwhandconformalmatter}, is unchanged
\begin{align}\label{eq:CFTunits}
    W_{CFT}(L,P| \phi_b)=\mathcal{N}_{\bar L,P| \phi_b}\text{exp}\left(\frac{\pi  c L  \sqrt{4 (\phi_b^2-P^2)+\frac{\pi ^2 c^2}{16}}}{ \left( \phi _b+P\right)}\right)\,,
\end{align}
but we have to remember that $\phi_b \sim P \sim M_p^2 R_{dS}^2$ so that the CFT term is subleading.
One can also redo the calculation of the prefactors, for example \eqref{eq:WignerPLbarvacuumwithprefactor} and one does not get any additional $M_p$, as we expect since it is a subleading one-loop effect:
\begin{align}
    \mathcal{N}_{L,P}=\pi L\,.
\end{align}
When adding matter further subleading effects will give corrections to the prefactor of order $c \, (M_p R_{dS})^{-2}$. 
For the distribution with the  inflaton introduced in section \ref{sec:additionofinflaton} we have
\begin{equation}\label{eq:inflatonunits}
W( L,Q,\Phi,P|f_b)= \mathcal{N}_{ L,Q,\Phi,P}\text{exp}\left(-\frac{\left[2\left(\Phi  L-P Q\right)+\pi  \lambda R_{dS}^4 ( L+Q) \left(2 f_b+\lambda \,R_{dS}^2 \right)\right]^2}{16 \lambda ^2 R_{dS}^6 \sqrt{Q^2- L^2}}\right)\,.
\end{equation}
Lastly, when we consider both the inflaton and the CFT, we get
\begin{align}
    &W(L,Q,\Phi,P|f_b) \nonumber\\
    &=\mathcal{N}_{ L,Q,\Phi,P}\exp\left(\frac{1}{4} \pi  c \sqrt{Q^2-L^2}-\frac{2  \left[2(\Phi L-P Q)+\pi  \lambda \, R_{dS}^4 (L+Q) \left(2 f_b+\lambda \, R_{dS}^2 \right)\right]^2}{\sqrt{Q^2-L^2} \left(\pi  c+32 \lambda^2 \, R_{dS}^6 \right)}\right).
\end{align}
\section{Entanglement Entropy}\label{app:entanglemententropy}
An analysis of the fine-grained entropy in various regions of the de Sitter spacetime can be found in references \cite{Chen:2020tes,Hartman:2020khs,Balasubramanian:2020xqf,Geng:2021wcq,Aalsma:2021bit,Kames-King:2021etp}.
It was argued in \cite{Chen:2020tes} that bra-ket wormholes must contribute to avoid an entropy paradox on the basic Hartle-Hawking background. In this appendix we study the resolution of the paradox on our geometry corresponding to the Wigner boundary conditions we outlined in the main text. The logic and the result is the same as in \cite{Chen:2020tes}, however, since the background solution is slightly different here, we find it useful to repeat the steps.
In our cosmological toy model, we consider glueing the expanding $dS$ region to flat space at the reheating surface as in figure \ref{fig:densitymatrixcontribution}.
The evolution in the gravitating region prepares a highly entangled state for the matter fields in the flat space region.
With such a state, we can compute the entropies of subregions using the gravitational fine-grained formula \cite{Penington:2019npb,Almheiri:2019psf}, which only requires the knowledge of the semiclassical geometry in the gravitating region. 
A thorough analysis of the entanglement entropy of subregions in this setting was performed in \cite{Chen:2020tes}, finding that the islands ensure that the entropy cannot grow larger than the $dS$ entropy, however the disconnected geometry on its own presents a violation of strong subadditivity.
In short, this happens when considering the entropy of intervals of length $l$ in the flat space region. The matter content consists of two CFTs: one with large central charge $c$ ($\text{CFT}_c$) for which the island entropy dominates and one ``probe'' CFT with small central charge $c_p$ ($\text{CFT}_p$) for which the no-island entropy dominates. One then considers the strong subadditivity inequality between the following subsystems:
\begin{align}\label{eq:SSA}
    S(A_c \cup A_p)+S(A_p \cup \overline{A}_p)-S(A_c \cup A_p \cup \overline{A}_p)-S(A_p)\geq 0\,,
\end{align}
where $A_{c,p}$ indicate the subsystems consisting of fields of the $\text{CFT}_{c,p}$ in the region $A$ which consists of an interval of length $l$ (or its complement $\overline{A}$) in flat space. This inequality gets violated when islands start appearing in the subsystems that include the $\text{CFT}_c$. This happens for large intervals:
\begin{align}\label{eq:HHislandsdominating}
    l \gg e^{\frac{6 \phi_0}{c}}\,.
\end{align}
In \cite{Chen:2020tes} it was shown that on bra-ket wormhole geometries the paradox is avoided. However, the latter had to be stabilized by adding some Euclidean evolution in the flat space region to become a solution. In this section, we study the entropy in the saddle we found through the Wigner distribution and we conclude that also in this setting the paradox is avoided. \par
\subsection{Entropy on bra-ket wormholes}
We found classical saddles for the bra-ket wormhole using the Wigner formalism. In general these have a complex geometry at the bra and ket boundaries, so when we glue to the non-gravitating region we have still have a complex geometry. It is somewhat subtle to define entanglement entropy in the Wigner distribution language since it is not a linear operator. However, we can consider a case where the backreaction is not too large and the contour endpoints are close to the original $\pi$-contour, which corresponds to $\ell_K=\ell_B$ and real. In this setting we can still use the islands formula to investigate the nature of the state using the wormhole as a classical background. 

Before reviewing the calculation of the entropy let us explain in detail what is the background that we consider and the approximations we use. The setup is the one introduced at the end of section \ref{sec:additionofinflaton} with the CFT with a large central charge $c\gg 1$. The inflaton can therefore be thought as a probe field solely to define a clock, and we can ignore its backreaction. The distribution and deviation from closed contour are given by \eqref{eq:LQFPwithcft} and \eqref{eq:teinflaton+CFT} in the limit of $\lambda \to 0$:
\begin{align}
    W_{\text{CFT}}(L,Q,\Phi,P|f_b)=&\mathcal{N}_{L,Q,\Phi,P}\exp\left(\frac{1}{4} \pi  c \sqrt{Q^2-L^2}-\frac{8 \left(\Phi L-P Q\right)^2}{\pi  c\sqrt{Q^2-L^2}}\right)\,,\\
    t_E=&-\frac{8 \left(\Phi L -P Q\right)}{\pi  c\sqrt{Q^2-L^2}}\,.
\end{align}
The distribution consists of two factors: the first pushes for large values of the Casimir energy and solely depends on $L,Q$, the second is a quadratic term that suppresses the phase space regions: 
\begin{align}
    \frac{8 \left(\Phi L-P Q\right)^2}{\pi  c\sqrt{Q^2-L^2}} \gtrsim 1\,,
\end{align}
giving typical values for $\Phi,P$ in the large Casimir energy regime. The condition of having an approximately  classical and continuous geometry within the spread of the distribution can be expressed via the parameter $t_E$ analogously to what is discussed in section \ref{subsec:interpretation_classicality}. It is (after rewriting in $U,V$ coordinates \eqref{eq:uvcoordinates}) 
\begin{equation}
    c L{\sqrt{V^2-1}} \gg 1\,.
\end{equation}
In the regime of $c \gg 1$, this bound is weaker than the one we have to put on the size of the Universe in the Euclidean regime to trust the effective description of the CFT given by Casimir energy and conformal anomaly \eqref{eq:CFTpartitionfunction}:
\begin{align}
   b \gg1 \iff  L\sqrt{V^2-1} \gg 1\,,
\end{align}
which, for large $L$, we can achieve with an appropriate inflaton potential as explained in footnote \ref{ref:inflationend}. We also focus on the regime for which the Universe is sufficiently large, such that $L \gg \frac{\Phi}{c}$, as in \cite{Chen:2020tes}, where the calculation of the islands entropy can be done in the OPE limit of twist operators.

Let us now move to the calculation of the entropy in our setup. The bra-ket wormhole solution is obtained in the Milne patch (we will stick to conformal coordinates here), which we first introduced in \eqref{eq:milnemetric} but repeat here for convenience
\begin{align}\label{eq:milnemetric2}
ds^2=\frac{-d\eta^2 + d\varphi^2 }{\sinh(\eta)^2}\,,\quad \phi=\frac{A}{-\tanh \eta}\,.
\end{align}
In these coordinates we have $\varphi \sim \varphi+2 \pi b$. Up to the conformal factor, this is the metric of Minkowski space. The relation between this and the flat space metric
\begin{align}\label{eq:flatspacemetric}
    ds^2= -dt_M^2+dx^2,
\end{align}
is fixed by glueing at the reheating surface $\eta=\eta_c$ to
\begin{align}
    x= \frac{\varphi}{- \sinh \eta_c}.
\end{align}
We have that $\varphi \sim \varphi+2 \pi b$ implies $x \sim x+2 \pi b/(-\sinh \eta_c)$, where $\eta_c$ is also fixed in terms of the Wigner boundary conditions. The state of the CFT produced in the non-gravitating region is now a thermal state at inverse temperature $\beta= i \Delta \eta$, more precisely, at the rescaled inverse temperature $\beta_x= (- \sinh{\eta_c} ) \beta$. This parameter is given in equation \eqref{eq:temperature} but since the contour is approximately closed we can neglect the corrections and stick to $\beta=\pi$.
We are working with a compact spatial slice but we imagine that the spatial circle is large, so that the entropy for a segment of length $l$ in the flat space region, i.e. $\Delta \varphi= l \, (-\sinh(\eta_c)) \simeq l (-\eta_c) $ in the bulk, is \cite{Calabrese_2004}:
\begin{align}\label{eq:noislandentropybraket}
    S_{\text{no-island}}=\frac{c}{3}\log\left(\frac { \sinh(\Delta \varphi )}{\varepsilon_{\text{uv}, \varphi}}\right)=\frac{c}{3}\log\left(\frac {(-\eta_c) \sinh(l/(-\eta_c))}{\varepsilon_{\text{uv}}}\right)\,,
\end{align}
where $\varepsilon_{\text{uv}}=(-\eta_c)\varepsilon_{\text{uv},\varphi}$. 
\begin{figure}
    \centering
    \subfigure[]{
    \label{fig:braketislands1}
    \includegraphics[width=0.23\textwidth]{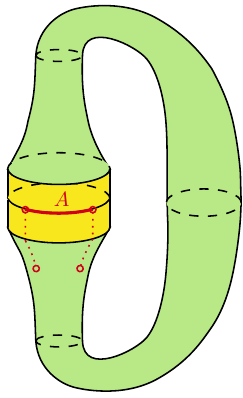}
    }
    \hspace{70pt}
    \subfigure[]{
    \label{fig:braketislands2}
    \includegraphics[width=0.23\textwidth]{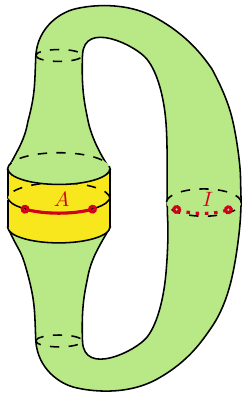}
    }
    \caption{a) The ansatz for islands in the gravitating region (green). The subregion $A$ is in the non-gravitating region (yellow), to emphasize this, we added in the figure some time evolution in the flat space region even though we computed the entropy right after glueing the two at the reheating surface. The position of twist operators is denoted by red circles, the cuts are represented by dotted lines. b) After finding an island in the throat (at $\eta= i \pi /2-\tilde \eta$), we can interpret the entropy computed through the gravitational fine grained entropy formula as the entropy of the subsystems $A\cup I$ entangled in the TFD state, since the island lies at a $\mathbb{Z}_2$ symmetric point of the torus.}
    \label{fig:braketislands}
\end{figure}
We can now look for islands. For this we need the explicit solution of the dilaton in the presence of conformal matter. As explained above, we ignore inflaton backreaction effects), such that we have:
\begin{equation}\label{eq:backreacteddilatonsolution}
    \phi = \frac{A}{-\tanh(\eta)} - \frac{c \pi^2}{3}\left(\frac{1}{\beta^2}-\frac{1}{4 \pi^2}\right)\left(\frac{\eta}{\tanh(\eta)}-1\right)\,,
\end{equation}
where we have explicitly included the dependence on the temperature induced by the non-contractible time cycle of the bra-ket wormhole, see figure \ref{fig:braketwormholewithparameters}, which will be set to $\beta=\pi$ below.
In the approximation $c\gg 1$, the search for extremal surfaces can be restricted to a configuration with the same spatial point as the end-points of the interval $l$ in the flat space region. This is justified in the OPE limit of the underlying twist operators. The ansatz for the two islands that lie at some generic point $\eta$ in the bulk at the same spatial point as the endpoints (depicted in figure \ref{fig:braketislands1}) is therefore (using equation \eqref{eq:backreacteddilatonsolution})
\begin{align}\label{eq:entropyansatzmilne}
    S_{\text{gen}}=2 \left\{\phi_0+ \frac{A}{-\tanh \eta}- \frac{c}{4}  \left[\frac{\eta}{\tanh \eta}-1\right]+\frac{c}{6}\log\left[ \frac{( \sinh(\eta))^2}{-\sinh(\eta) \varepsilon_{\text{uv}, \varphi}}\right]\right\}.
\end{align}
From the result of \cite{Chen:2020tes} we expect the islands to be in the middle of the contour in Milne coordinates i.e. in the throat of the wormhole. This point can be parameterized by $\eta= i \pi/2-\tilde \eta$, the ansatz becomes
\begin{align}\label{eq:genentropythroat}
    S_{\text{gen}}=2 \left\{\phi_0+ A \tanh\tilde\eta+ \frac{c}{4} \left[\left(\frac{i \pi}{2}-\tilde\eta\right) \tanh \tilde \eta +1\right]+\frac{c}{6}\log\left[-i \frac{\cosh \tilde \eta}{\varepsilon_{\text{uv}, \varphi}}\right]\right\}.
\end{align}
To ensure that the $\Phi$ argument of the Wigner distribition is real, we have to give an imaginary part to $A$: $A=A_r -i \frac{\pi c}{8}$, and we see that this cancels the imaginary part of the dilaton term in the entropy. The on-shell value of $A_r$ is (in small $t_E$ limit justified by classicality):
\begin{align}\label{eq:ArCFTsmallclargetc}
   A_r= \frac{\Phi\sqrt{Q^2-L^2}}{Q}\,.
\end{align}
The phase in the argument of the logarithm does not change the extremization procedure and we discard it, getting to the following equation for the extremum
\begin{align}\label{eq:extremumthroat}
    -24 A_r+6 c \tilde \eta +c \sinh (2 \tilde \eta )=0.
\end{align}
This equation always has a real solution, meaning that there is an island in the throat of the wormhole. However, the fact that \eqref{eq:extremumthroat} is a trascendental equation forces us to consider some limits to get analytic expressions. In the large $c$ regime we can consider $\tilde \eta$ to be small. We get the following extremum
\begin{align}
    \tilde\eta^*= \frac{3 A_r}{c}.
\end{align}
The entropy is then obtained by plugging this in the equation \eqref{eq:genentropythroat} and discarding a phase inside the logarithm that does not affect the extremization procedure, such that we get to \footnote{We also average over the position of the islands that can be either at the bra or at the ket, in the $c \gg 1$ limit doing this just cancels the imaginary term in the ansatz.}
\begin{align}\label{eq:braketentropysmalleta}
   S= 2\left[\phi_0+\frac{c}{4}+\frac{1}{6} c \log \left(\frac{1}{\varepsilon_{\text{uv}, \varphi} }\right)+\mathcal{O}\left(\frac{A_r^2}{c}\right)\right].
\end{align}
Notice that in this limit the entropy does not depend on $A_{r}$ and therefore does not depend on the amount of Lorentzian evolution in the de Sitter region. By comparing with the no-island entropy \eqref{eq:noislandentropybraket} we see that the former dominates if
\begin{align}
    \Delta \varphi \gg 6 \frac{\phi_0}{c} \,, \quad l  \gg (-\eta_c) 6 \frac{\phi_0}{c}\,.
\end{align}
Notice that these islands appear for shorter intervals than the Hartle-Hawking ones \eqref{eq:HHislandsdominating}.
As we said above, the extremum equation \eqref{eq:extremumthroat} always has a solution in the throat of the wormhole. In that region, the metric as a function of $\tilde \eta$ is:
 \begin{align} \label{eq:throatgeometry}
     \phi(\tilde \eta)&= A \tanh \tilde \eta +\frac{c}{4}\left(1-\tilde \eta \tanh \tilde \eta\right),\\
     ds^2&=\frac{d \tilde \eta ^2-d \varphi^2}{\cosh \tilde \eta^2}.
 \end{align}
It would be interesting to see if this region of spacetime, where the island can be located, has any physical meaning. For the double-cone geometry an analogous region was discussed in \cite{Chakravarty:2024bna}.
 The fact that we find an island in the throat allows us to interpret the entropy obtained through the islands formula as the entropy of the CFT fields in the flat space region entangled with another system which is comprised of the CFT and geometry in the throat \eqref{eq:throatgeometry}, in the TFD state $|A \cup I\rangle_{\text{TFD}}$, as in figure \ref{fig:braketislands2}. As in \cite{Chen:2020tes} this allows to resolve the subadditivity paradox found using the Hartle-Hawking geometry. Let us report the argument here. The inequality to consider is \eqref{eq:SSA} and the islands configuration for the calculation of the entropy of the various subregions involved is summarized in figure \ref{fig:paradoxsolved}.
\begin{figure}[]
    \centering
    \subfigure[]{
\includegraphics[width=0.4\textwidth]{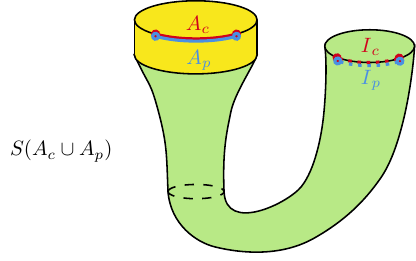}
    }
    \subfigure[]{
    \hspace{10pt}
\includegraphics[width=0.4\textwidth]{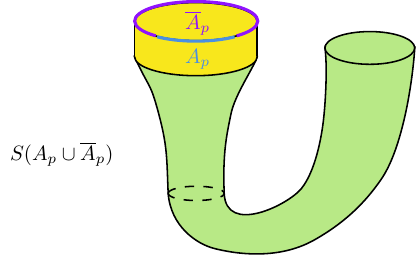}
    }
    \hspace{40pt}
    \subfigure[]{
\includegraphics[width=0.4\textwidth]{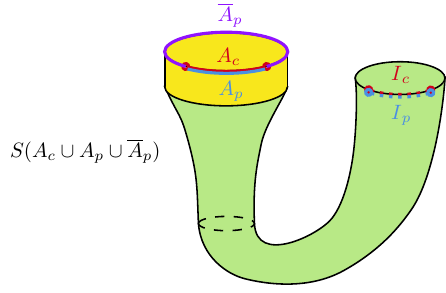}
    }
     \hspace{35pt}
    \subfigure[]{
\includegraphics[width=0.37\textwidth]{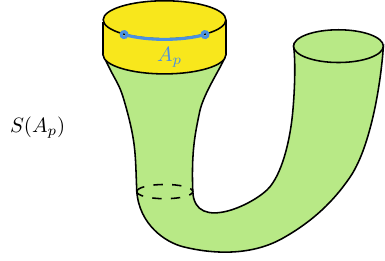}
    }
    \caption{Islands configuration for the calculation of the entropy of the various subregions involved in the SSA inequality \eqref{eq:SSA}. We draw only the half-wormhole between the flat space region and the throat. In a) and c) the islands appear because of the large central charge $c$, it is important to notice that the additional twist operators added following the QES prescription do not distinguish between the two CFTs so one has islands for both the CFTs: $I_c$ and $I_p$. In b) and d) instead the no-island entropy dominates.}
    \label{fig:paradoxsolved}
\end{figure}
For the contribution for which the islands dominate we have
\begin{align}
	S(A_{p}\cup A_{c})&= \text{Area terms}+S_{\text{bulk}}(A_{c}\cup I_{c})+S_{\text{bulk}}(A_{p}\cup I_{p})\,,\\
	S(A_{p} \cup \overline{A}_{p} \cup A_{c})&= \text{Area terms}+S_{\text{bulk}}(A_{c}\cup I_{c})+S_{\text{bulk}}(A_{p}\cup  \overline{A}_{p} \cup I_{p})\,.
\end{align}
Whereas for the configurations for which the no-islands entropy dominates we have
\begin{align}
	S(A_{p} \cup \overline{A}_{p})&= S_\text{bulk}(A_{p} \cup \overline{A}_{p})\,,\\
	S(A_{p})&= S_\text{bulk}(A_{p})\,.
\end{align}
Putting all together, the area terms as well as the entropy of $A_{c}\cup I_{c}$ cancel and we remain with
\begin{align}
    S_{\text{bulk}}(A_p \cup I_{p})+S_{\text{bulk}}(A_p \cup \overline{A}_p)-S_{\text{bulk}}(A_p \cup \overline{A}_p \cup I_{p})-S_{\text{bulk}}(A_p)\geq 0.
\end{align}
Which is satisfied because it is a strong subadditivity inequality for the $\text{CFT}_{p}$ in the TFD state.
\subsection{Entropy on Hartle-Hawking: checking that the paradox never appears}
To check that the paradox does not appear, we will argue that whenever the problematic island in the disconnected geometry appears, the wormhole dominates. To do so, we first have to evaluate the entropy for the global Hartle-Hawking geometry. As we have done for the connected solution, we stick to a configuration where the geometry is continuous, this corresponds to the situation where we consider the point in phase space that corresponds to the peak of the delta functions \eqref{eq:HHLQPWigner}. The relevant metric is \eqref{eq:globalmetric}, in general there are two different integration constants for the dilaton on the bra and ket branches of the contour, which in the limit where the geometry is continuous coincide and are equal to
\begin{align}
  A_{G,K}=A_{G,B}= \frac{\Phi}{L}\equiv A_G\,.
\end{align}
In this case, the spatial circle is compact and of length $\tilde L =2 \pi L$. Where $L$ is the argument of the Wigner distribution (the average of \eqref{eq:boundarylengthdefinition} at bra and ket). The Hartle-Hawking state corresponds to the vacuum for the CFT, therefore the no-islands entropy for an interval is the standard one for a compact spatial slice \cite{Calabrese_2004}:
\begin{equation}
    S_\text{no-island}= \frac{c}{3}\log\left(\frac{2 \sin\left(\frac{\Delta \hat \varphi}{2}\right)}{\varepsilon_{\text{uv},\hat \varphi}}\right)=\frac{c}{3}\log\left(\frac{\tilde{L} \sin\left( \frac{\pi l}{\tilde L}\right)}{ \pi \varepsilon_{\text{uv}}}\right)\,,\quad \varepsilon_{\text{uv},\hat \varphi}= \frac{2 \pi}{\tilde L} \varepsilon_{\text{uv}}\,.
\end{equation}
Notice that this entropy is maximized at $l=\tilde L /2$. We can also look for islands, as we have done before, we search for islands in the bulk at the same spatial point as the end-point of the interval in the flat space region, see figure \ref{fig:HHislands}.
\begin{figure}
    \centering
    \includegraphics[width=0.15\textwidth]{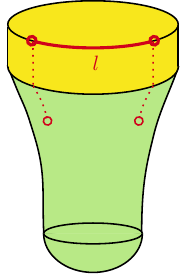}
    \caption{Depiction of the ansatz for the generalized entropy to determine the entropy of an interval after preparing the state in the Hartle-Hawking geometry.}
    \label{fig:HHislands}
\end{figure}

Therefore, the ansatz for the generalized entropy is
\begin{align}
    S_\text{gen}= 2 \left[\phi_0+ \frac{A_G}{-\sin(\hat{\eta})}+ \frac{c}{6}\log \left(\frac{\left(2 \sin \left(\frac{ \hat\eta }{2}\right)\right)^2}{-  \sin (\hat \eta )\varepsilon_{\text{uv},\hat{\varphi}}}\right)\right]+\frac{i c \pi}{6}\,,
\end{align}
Extremizing this gives the following result 
\begin{align}
    \sin(\hat{\eta}^*)=- \frac{6 A_G}{c}\,.
\end{align}
We are interested in the regime of $L \gg \Phi/c$ which implies $A_G/c \ll 1$, in this limit the entropy is given by
\begin{align}
    S_\text{island}=2 \phi_0+ \frac{c}{3}+\frac{c}{3}\log\left(\frac{6 A_G}{c\, \varepsilon_{\text{uv},\hat{\varphi}}}\right)+ \mathcal{O}\left(A_G^2/c\right)\,.
\end{align}
In order to have an island we need that:
\begin{align}
    S_\text{island}-S_\text{no-island}=2 \phi_0+ \frac{c}{3}+\frac{c}{3}\log\left(\frac{3\Phi}{c} \frac{1}{L \sin \left(\frac{l}{2L}\right)}\right)\,\leq 0\,,
\end{align}
where we used that in the Hartle-Hawking case $A_G=\Phi/L$ and substituted spatial length of the circle $\tilde L$ with the argument of the Wigner distribution $\tilde L=2 \pi L$. For very large spatial circles, corresponding to $ L\gg l$ we recover the results of \cite{Chen:2020tes} including the value of $l$ for which the island starts dominating (eq. \eqref{eq:HHislandsdominating}) and strong subadditivity is violated. Of course also in the connected case one finds the same paradox for large enough intervals. However, from this calculation we see that in order for this island to ever appear there is a lower bound on $L$:
\begin{align}\label{eq:Lprobislands}
     L \geq L_\text{prob}\equiv \frac{3 \Phi}{c}e^{\frac{6\phi_0}{c}+1}\,.
\end{align}
Since in the wormhole the paradox is avoided, we just have to check that whenever this problematic island can appear (i.e. $L\geq L_\text{prob}$) the wormhole dominates. We will argue now that this is the case in the approximations we are working in.

The comparison between the Wigner distribution of Hartle-Hawking and wormhole is analogous to what we discussed in section \ref{subsec:comparisonHHbraket}, with the difference that the bra-ket wormhole is further enhanced by the Casimir term:
\begin{align}
\frac{W_{\text{bra-ket}}}{W_{\text{HH}}}\approx \frac{L^4}{\Phi^2} e^{-2\phi_0+\frac{1}{4} \pi  c L \sqrt{V^2-1}}\,.
\end{align}
Let us neglect the prefactors, which further suppress the disconnected solution for large universes. The wormhole dominates for
\begin{align}
    L>L_c \equiv \frac{6 \phi_0}{c} \frac{4}{3 \pi}\frac{1}{\sqrt{V^2-1}}\,.
\end{align}
Comparing with \eqref{eq:Lprobislands} we can see that the threshold for the problematic island to ever appear is much larger than the critical size of the universe for which the wormhole dominates, i.e.
\begin{align}
    L_{\text{prob}} \gg L_c \iff  \frac{3 \Phi}{c}e^{\frac{6\phi_0}{c}+1}\gg \frac{6 \phi_0}{c} \frac{4}{3 \pi} \frac{1}{\sqrt{V^2-1}}\,
\end{align}
which is guaranteed for any value of $\phi_0/c$ in the regime we are working in via the bounds
\begin{align}
    \frac{1}{\sqrt{V^2-1}}\ll L \ll \frac{\Phi}{c}\,.
\end{align}
\section{Wigner Distribution} \label{sec:app_wigner}
\subsection{Phase space quantum mechanics}
In this section we review the basic definitions of the phase space formulation of quantum mechanics, see for example \cite{Polkovnikov_2010} for an extended review. Here we will focus on the Wigner-Weyl representation and stick to a coordinate-momentum basis which is directly generalizable to our cosmology minisuperspace system. As opposed to the standard Schr\"{o}dinger picture it allows us to consider quantities as functions of coordinates and momenta, instead of working in one or the other basis. The formalism builds upon the standard operators in quantum mechanics $\hat{A}$, for which one defines the Wigner transform as follows
\begin{align} 
    A_W(X,P)= \int d \Delta x  \, \left\langle X + \frac{\Delta x}{2} \left\rvert {\hat{A}}\right\lvert{X-\frac{\Delta x}{2}}\right\rangle e^{i \frac{P \cdot \Delta x }{ \hbar}}\,,
\end{align}
and the Wigner distribution, i.e. the Wigner transform of the density matrix as 
\begin{equation}
    W(X,P;t)=\int d \Delta x  \, \rho\left( X + \frac{\Delta x}{2}, X - \frac{\Delta x}{2}|t\right)\, e^{i \frac{P \cdot \Delta x }{  \hbar}}.
\end{equation}
For several variables there is a straightforward generalization, which also allows to consider ``mixed'' objects:
\begin{equation}
    W(X,P;y_B,y_K|t)=\int d \Delta x  \, \rho\left( X + \frac{\Delta x}{2}, X - \frac{\Delta x}{2};y_B,y_K|t\right)\, e^{i \frac{P \cdot \Delta x }{  \hbar}}.
\end{equation}
If the state is a pure state $\rho=\ket{\psi}\bra{\psi}$ then
\begin{equation}
    W(X,P|t)=\int d \Delta x  \; \psi\left(X +\frac{\Delta x}{2},t\right) \psi^*\left(X  - \frac{\Delta x}{2},t\right)\,e^{i \frac{P \cdot \Delta x }{ \hbar}}.
\end{equation}
Given the two elements above, one can write expectation values of operators $\hat{A}$ on a state characterized by $W(x,p)$ as
\begin{align}\label{eq:expvalA}
    \langle\hat{A}\rangle= \int dP\, dX\, W(X,P) A_W(X,P)\,.
\end{align}
It can be easily checked that this gives the usual $\text{Tr}(\hat{\rho}\hat{A})$. We can also derive an equation for the time evolution of the Wigner distribution, by Wigner transforming the evolution equation for the density matrix as one gets
\begin{align}\label{eq:wigner_evol}
    \partial_t{W}(X,P|t)= -\left\{H_W,W \right\}_M\,.
\end{align}
Here $\left\{A_W,B_W\right\}_M$ is the Moyal bracket between the two Wigner transformed operators, defined as
\begin{align}
    \left\{A_W,B_W\right\}_M=- \frac{2}{\hbar} A_W \sin\left(\frac{2}{\hbar} \Lambda\right)B_W\,,
\end{align}
and
\begin{align}
    \Lambda= \overset{\leftarrow}{\frac{\partial}{\partial P}}\overset{\rightarrow}{\frac{\partial}{\partial X}}-\overset{\leftarrow}{\frac{\partial}{\partial X}}\overset{\rightarrow}{\frac{\partial}{\partial P}}\,,
\end{align}
where the arrows indicate what the derivatives act upon. The Moyal brackets reduce to ordinary Poisson brackets in the semiclassical limit, and equation \eqref{eq:wigner_evol} reduces to the ordinary Liouville equation for a classical phase space probability distribution:
\begin{align}\label{eq:wigner_evol_semicl}
    \frac{\partial W}{\partial t}(X,P|t)= -\left\{H_W,W \right\}_M= -\left\{H_W,W \right\}_\text{Poisson} + \hbar^2 \displaystyle \sum_{i,j,k} \frac{\partial^3 V}{\partial X_i \partial X_j \partial X_k}\frac{\partial^3 W}{\partial P_i \partial P_j \partial P_k}+\dots \,.
\end{align}
Up to $\mathcal{O}(\hbar^2)$ (or considering quadratic potentials) the above equation reduces to the classical Liouville equation for a phase space distribution. This motivates us to look again at equation \eqref{eq:expvalA}, and notice that in the classical limit we can interpret $W$ as a probability distribution and $A_W$ as the classical observable function of coordinate and momenta. However, as it is not in general guaranteed that the Wigner distribution is positive definite, one must be careful with this interpretation. 
\subsection{Semiclassical features: harmonic oscillator}
To see how the Wigner distribution behaves for semiclassical states we can take the simplest example, the harmonic oscillator. We use units in which all the parameters of the harmonic oscillator and $\hbar$ have unit value. The most direct point of contact with the classical phase space distribution arises when considering a thermal state at inverse temperature $\beta$. The Wigner function can be computed using, for example, Euclidean path integrals. The density matrix is the Euclidean propagator on a circle of length $\beta$
\begin{align}
    \rho(x_K,x_B|t=-i\beta)= \mathcal{N} \exp\left(-\frac{1 }{2} \frac{(x_K^2+x_B^2)\cosh \beta-2 x_K x_B}{\sinh  \beta}\right).
\end{align} 
After properly normalizing the density matrix to $\text{Tr}(\rho)=1$ and doing the Wigner transform, we get
\begin{align}
     W(X,P|\beta) = 2 \tanh\left(\frac{\beta}{2}\right)\exp\left[-4\left(P^2+X^2  \right) \tanh( \beta/2)\right].
\end{align}
For low temperatures it reduces to the Wigner distribution of the ground state \cite{Polkovnikov_2010}:
\begin{align}
    W(X,P|\beta \to \infty)=\exp\left[-\left(P^2+X^2  \right)\right].
\end{align}
Instead, for high temperatures we get
\begin{align}\label{eq:harmonicoscillatorboltzmannwigner}
    W(X,P|\beta\to 0)= \beta \exp\left[- \frac{\beta}{2}\left( {P^2}+X^2\right)\right],
\end{align}
which is the classical Boltzmann distribution of the harmonic oscillator. So that we see that the Wigner distribution for a semiclassical state (high temperature) reduces to a classical phase space distribution.

Moreover, for generic semiclassical states one can show using the WKB approximation that the Wigner distribution is peaked at classical trajectories in phase space. The WKB wave function has the form
\begin{align}
    \psi(x)=C(x,t) e^{i S(x,t)}.
\end{align}
Computing the Wigner distribution in the semiclassical limit, one gets (see for example \cite{Halliwell:1987eu})
\begin{align}
    W_{\text{WKB}}(X,P|t)=\lvert{C(X,t)}\lvert^2\delta\left(P-\frac{\partial S(X,t)}{\partial X}\right)\,,
\end{align}
which is peaked at trajectories in phase space that satisfy the Hamilton-Jacobi equation 
\begin{align}
    P-\frac{\partial S(X,t)}{\partial X}=0.
\end{align}
Let us now see how we recover this in the simple example of the harmonic oscillator.
We quote an explicit example from reference \cite{Halliwell:1987eu}, and then compare to our bra-ket wormhole Wigner distribution. Consider again the simple harmonic oscillator. One can compute the wave function depending on a parameter $\phi$ that labels the states
\begin{align}
    \psi_\phi(x,t)=A(t)e^{-B(t)x^2}\,,
\end{align}
with
\begin{align}
    A(t)&=\left[\frac{ \sin(2 \phi)}{2\pi  \cos^2(\phi+t)}\right]^{1/4}\\
    B(t)&= \frac{\sinh(2 \phi)-i \sin(2 t )}{\cosh(2 \phi)- \cos(2 t)}\,.
\end{align}
For $\phi\to \infty$ the wave function becomes the one of the ground state of the harmonic oscillator, and the resulting Wigner function describes this ground state.  For small $\phi$ (corresponding to excited states) the Wigner function is
\begin{align}
    W_{\phi \to 0}(X,P|t)=\exp\left(-\frac{2 \sin^2(t)}{ \sinh \phi}\left[P - X\cot^2(t)\right]^2\right).
\end{align}
Which is peaked at 
\begin{align}
    P= X \cot^2(t),
\end{align}
This is a first integral of the equations of motion for the harmonic oscillator with momentum $P$ and position $X$, i.e. for a semiclassical system the Wigner distribution is peaked at classical trajectories in phase space. In fact this is what we find for the Wigner distribution with the inflaton, formula \eqref{eq:LFQPdistribution}, as it features a peak in phase space within the approximation used.

\bibliography{main}
\bibliographystyle{utphys.bst}

\end{document}